\documentclass[aps,                
               prc,                
               showpacs,           
               showkeys,           
               twocolumn,          
               10pt,               
               twoside,            
               floatfix,           
               superscriptaddress] 
               {revtex4-1}

\usepackage{mathptmx}         
\usepackage[T1]{fontenc}
\usepackage{geometry}         
\usepackage{fancyhdr}         
\usepackage{graphicx}         
\usepackage[svgnames]{xcolor} 
\usepackage{hyperref}         
\usepackage{amsmath}          
\usepackage{amssymb}          
\usepackage{amsfonts}         
\usepackage{amstext}          
\usepackage{mathrsfs}         
\usepackage{array}            
\usepackage{textcomp}         

\hypersetup{colorlinks=true,linkcolor=Blue,citecolor=Blue,urlcolor=Blue}

\DeclareGraphicsExtensions{.pdf}

\fancypagestyle{firststyle}
{%
  \fancyhf{}
  \fancyhead[C]{PHYSICAL REVIEW C {\bf 92}, 015201 (2015)} 
  \fancyfoot[L]{\small 0556-2813/2015/92(1)/015201(14)} 
  \fancyfoot[C]{\small 015201-\thepage} 
  \fancyfoot[R]{\small \textcopyright 2015 American Physical Society} 
}

\newcolumntype{I}{!{\vrule width 1.1pt}}

\newlength{\savedwidth}

\newcommand{\whline}%
{%
  \noalign{\global\setlength{\savedwidth}{\arrayrulewidth}}%
  \noalign{\global\setlength{\arrayrulewidth}{1.1pt}}\hline%
  \noalign{\global\setlength{\arrayrulewidth}{\savedwidth}}%
}

\newcommand{\wcline}[1]%
{%
  \noalign{\global\setlength{\savedwidth}{\arrayrulewidth}}%
  \noalign{\global\setlength{\arrayrulewidth}{1.1pt}}\cline{#1}%
  \noalign{\global\setlength{\arrayrulewidth}{\savedwidth}}%
}

\begin{document}

  \title{\texorpdfstring{Observables in ultrarelativistic heavy-ion collisions from two different transport\\approaches for the same initial conditions}{Observables in ultrarelativistic heavy-ion collisions from two different transport approaches for the same initial conditions}}

\author{R.~Marty}
\email[Email : ]{marty@fias.uni-frankfurt.de}
\affiliation{Frankfurt Institute for Advanced Studies, Johann Wolfgang Goethe Universit\"at, Ruth-Moufang-Strasse 1,\\ 60438 Frankfurt am Main, Germany}
\affiliation{\begin{minipage}[c]{0.93\textwidth}Institute for Theoretical Physics, Johann Wolfgang Goethe Universit\"at, Max-von-Laue-Strasse 1, 60438 Frankfurt am Main, Germany\end{minipage}}

\author{E.~Bratkovskaya}
\affiliation{Frankfurt Institute for Advanced Studies, Johann Wolfgang Goethe Universit\"at, Ruth-Moufang-Strasse 1,\\ 60438 Frankfurt am Main, Germany}
\affiliation{\begin{minipage}[c]{0.93\textwidth}Institute for Theoretical Physics, Johann Wolfgang Goethe Universit\"at, Max-von-Laue-Strasse 1, 60438 Frankfurt am Main, Germany\end{minipage}}

\author{W.~Cassing}
\affiliation{\begin{minipage}[c]{0.73\textwidth}Institut f\"ur Theoretische Physik, Universit\"at Gie\ss{}en, Heinrich-Buff-Ring 16, 35392 Gie\ss{}en, Germany\end{minipage}}

\author{J.~Aichelin}
\affiliation{\begin{minipage}[c]{0.98\textwidth}Subatech, UMR 6457, IN2P3/CNRS, Universit\'e de Nantes, \'Ecole des Mines de Nantes, 4 rue Alfred Kastler, 44307 Nantes cedex 3, France\end{minipage}\\
\normalfont{(Received 17 December 2014; revised 3 April 2015; published 7 July 2015)}\vspace{2.5mm}
}

\pacs{25.75.Nq, 24.10.Lx, 12.39.Ki,51.30.+i%
\hfill DOI:\href{http://dx.doi.org/10.1103/PhysRevC.92.015201}{10.1103/PhysRevC.92.015201}}



\begin{abstract}
For nucleus-nucleus collisions at energies currently available at the BNL Relativistic Heavy Ion Collider (RHIC), we calculate observables in two different transport approaches,
i.e., the $n$-body molecular dynamical model ``relativistic quantum molecular dynamics for strongly interacting matter with phase transition or crossover'' (RSP) and the two-body
parton hadron string dynamics
(PHSD), starting out from the same distribution in the initial energy density
at the quark gluon plasma (QGP) formation time. The RSP dynamics is based
on the Nambu-Jona-Lasinio (NJL) Lagrangian, whereas in PHSD the partons
are described by the dynamical quasiparticle model (DQPM). Despite the
very different description of the parton properties and their interactions
and of the hadronization in both approaches, the final transverse momentum
distributions of pions turn out to be quite similar, which is less visible
for the strange mesons owing to the large NJL cross sections involved.
Our findings can be attributed, in part,
to a partial thermalization of the  quark degrees of freedom in central Au+Au
collisions for both approaches.
The rapidity distribution of  mesons shows a stronger sensitivity to
the nature of the degrees of freedom involved and to their
interaction strength in the QGP.
\end{abstract}

\maketitle

  \section{Introduction}
\thispagestyle{firststyle}
The primary objective of the study of ultrarelativistic heavy-ion collisions at the BNL Relativistic Heavy Ion Collider (RHIC) and at the CERN Large Hadron Collider (LHC) is to search for a new state of matter, a plasma of quarks and gluons (QGP)  predicted at high temperatures and high densities  \cite{Bohr1977}.  Such a state presumably existed in the primordial universe shortly after the big bang. In the QGP the fundamental particles of the strong interaction --the quarks and the gluons-- are deconfined. At lower temperatures and densities the quarks and gluons are confined in hadrons, the particles that can be observed in experiments.

Lattice gauge calculations \cite{Aoki2005} predict that  for a static infinite medium in equilibrium and at zero chemical quark potential $\mu $,  the transition from  hadronic matter to the QGP (and vice versa) is a cross over. In heavy-ion collisions this new state of matter can only be produced in a finite interaction zone of a radius of less than 10 fm and its lifetime is, at most, of the order of 8-10 fm/$c$ ($\sim 3 \times 10^{-23}$ s). In addition, up to now the short equilibration time of the order of 1 fm/$c$, necessary to form a QGP, is not yet explained convincingly by theory. Therefore, the study of the formation of a QGP in heavy-ion collisions raises several issues.

First, the finite value of the third coefficient in the Fourier expansion of the azimuthal hadron angular distribution $v_3$ shows that the form of the interaction zone fluctuates from event to event considerably even for reactions at a fixed impact parameter. One may consider the initial stage either as an evolution of color fields \cite{Albacete2014} or as a string formation and subsequent string melting \cite{Andersson1983,Andersson1998}. Despite a completely different physical origin of the fluctuations, both approaches can give a reasonable agreement with experiment, however, with different sets of parameters (see Refs. \cite{Luzum2008,Gale2013} and references therein). Therefore, the physically correct description of the heavy-ion reaction in the first fm/$c$ is still debated at this time \cite{Konchakovski2014}.

Second, it is still an open question if in these reactions after $\sim$ 1 fm/$c$ a local thermal equilibrium is attained. There is no theoretical guidance yet as to how this may happen and therefore the energy-density distribution and the local velocity fields at the beginning of the expansion of the QGP are afflicted with a lot of uncertainties.

Approaches which that the expansion of the QGP in hydrodynamical models assume a local thermal equilibrium and are able to describe a multitude of observables. This raises the question of whether the agreement with experiment is sufficient to justify the assumption of a local equilibrium at the beginning of the expansion. Other transport approaches such as \emph{a multiphase transport} (AMPT) model \cite{Lin2005} and the \emph{parton-hadron-string dynamics} (PHSD) \cite{Cassing2009,Bratkovskaya2011a} are based on Boltzmann or Kadanoff-Baym-type transport equations which do not require the assumption of a local equilibrium. There the initial conditions (using {\sc HIJING} \cite{Wang1991} for AMPT and {\sc FRITIOF} \cite{Andersson1998} + {\sc PYTHIA} \cite{Sjostrand2006} for PHSD) are directly converted into dynamical partons which subsequently interact by potential and collisional interactions.

Recently the \emph{relativistic quantum molecular dynamics for strongly interacting matter with phase transition or cross\-over} (RSP) \cite{Marty2012} has been advanced, which is a relativistic molecular-dynamics approach based on the Nambu-Jona-Lasinio (NJL) Lagrangian \cite{Nambu1961,Klevansky1992}. Being a $n$-body approach, it conserves the fluctuations of the initial conditions, an important requirement if one wants to study the first-order phase transition at higher chemical potential $\mu$, as predicted within the NJL.

Despite of the fact that all these approaches (more or less) describe the lattice equation of state, the properties of the partons are rather different in these models (cf. \cite{Marty2013}). In the dynamical quasiparticle model (DQPM) the masses of the partons are large at high temperature and close to $T_c$ and therefore the hadronization proceeds by the formation of large mass hadrons, which decay subsequently to the pseudoscalar octet. In the NJL approach gluons do not appear as explicit (timelike) degrees of freedom, but they appear as a potential interaction among quarks as spacelike degrees of freedom. At high temperature the hadrons have their bare mass in the NJL. Close to $T_c$ their mass increases owing to the increasing scalar condensate but remains small compared to the mass of quarks in the DQPM. The hadronization takes place via $q+\bar q \to $ hadron+hadron interactions for which the cross section becomes very large close to $T_c$ and below.

In view of the different degrees of freedom to realize a lattice equation of state in transport theories, it is important to know whether these approaches lead --for the same initial conditions-- to different values of the observables or whether observables can be identified that allow to fix the parton properties in the QGP.

The initial conditions of the different approaches, i. e., how to transform the initial stage of the heavy-ion collisions into a QGP plasma, described either by hydrodynamics or by transport-type equations, are quite different. Here we choose as the basis for the comparison the initial condition of PHSD, i.e., the initial spatial distribution of the energy density from the PHSD approach. We show how to transform the PHSD initial condition into the degrees of freedom of the RSP approach. This transformation is not unique and therefore we present several possibilities to perform this transformation. Having the same energy-density distribution we propagate the degrees of freedom using either the PHSD or the RSP equations. We discuss the initial spectra of partons, the final spectra of identified particles in both approaches, as well as the elliptic flow. Therefore, the purpose of this study is to demonstrate how the same initial energy-density distribution evolves in the two different approaches to the final hadron spectra. We recall that with the initial conditions adopted here the PHSD approach describes a multitude of experimental observables \cite{Toneev2012,Konchakovski2012,Konchakovski2012a,Linnyk2012,Linnyk2011,Linnyk2013,Linnyk2014,Linnyk2013a}. Therefore, by adapting the same initial conditions for the RSP we can investigate the sensitivity of the ``bulk'' observables to the initial conditions, the QGP dynamics, and the details of the hadronization.

In the next section we briefly present the PHSD approach and we discuss the initial energy-density profile  extracted. Then in Sect. III, we briefly present the NJL model and in Sect. IV we present the RSP transport approach. In Sect. V we present the comparison of both models for smooth initial conditions. In Sect. VI we explain how we convert these initial conditions into a plasma composed of NJL particles, with and without the assumption of thermal equilibrium, and describe in detail the case of an out-of-equilibrium conversion and its consequences. In Sect. VII we focus on the comparison of results from PHSD and RSP for the initial quark distributions and then for the final hadron spectra. We conclude our study with a summary in Sect. VIII. In the Appendix we discuss more about the hadronization.

\vspace*{-4mm}
\section{Realistic event-by-event initial conditions}
\vspace*{-2mm}

\begin{figure*}
  \begin{center}
    \includegraphics[width=8.5cm]{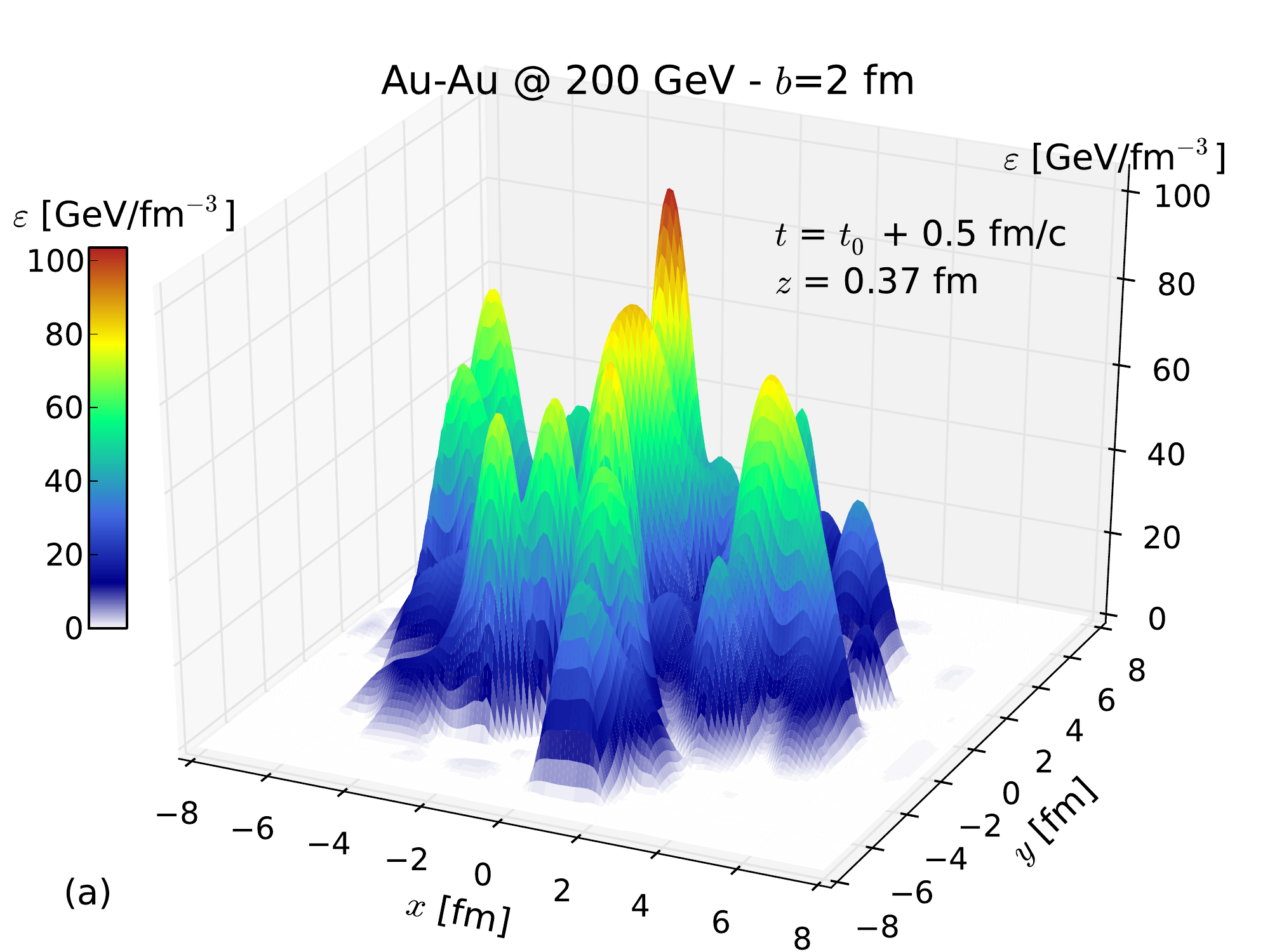} \ \ \ \
    \includegraphics[width=8.5cm]{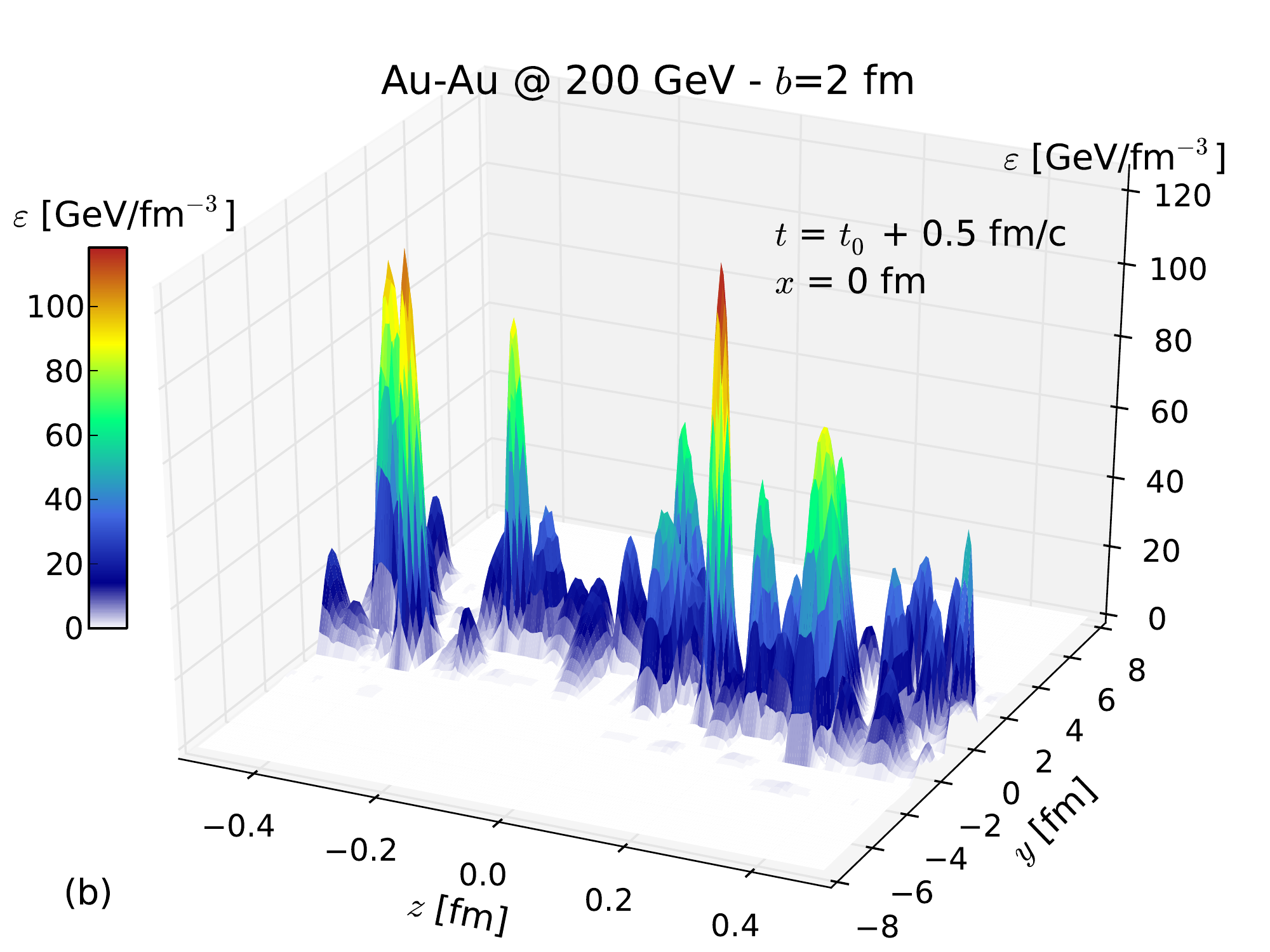}\\[2mm]
    \includegraphics[width=8.5cm]{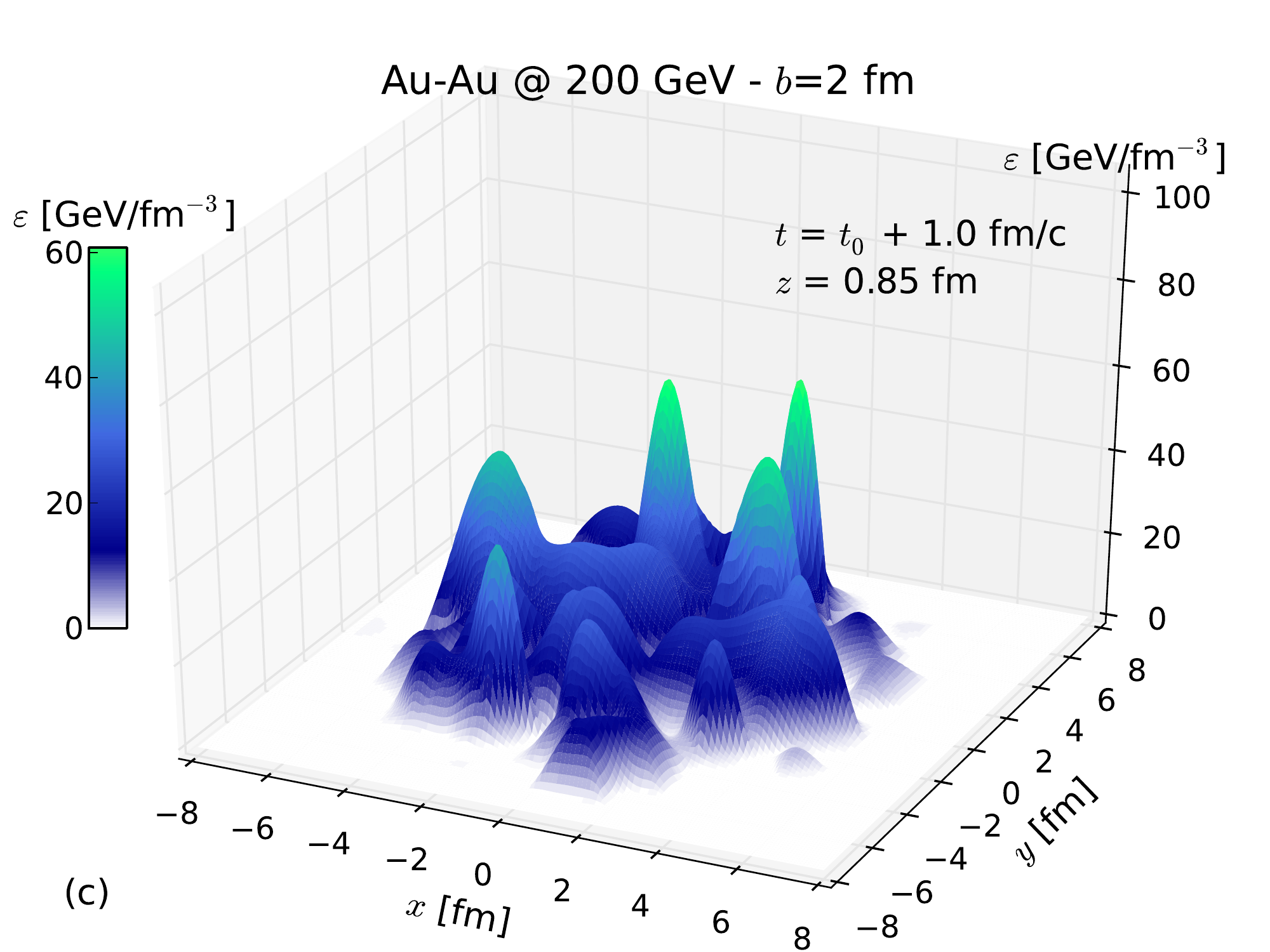} \ \ \ \
    \includegraphics[width=8.5cm]{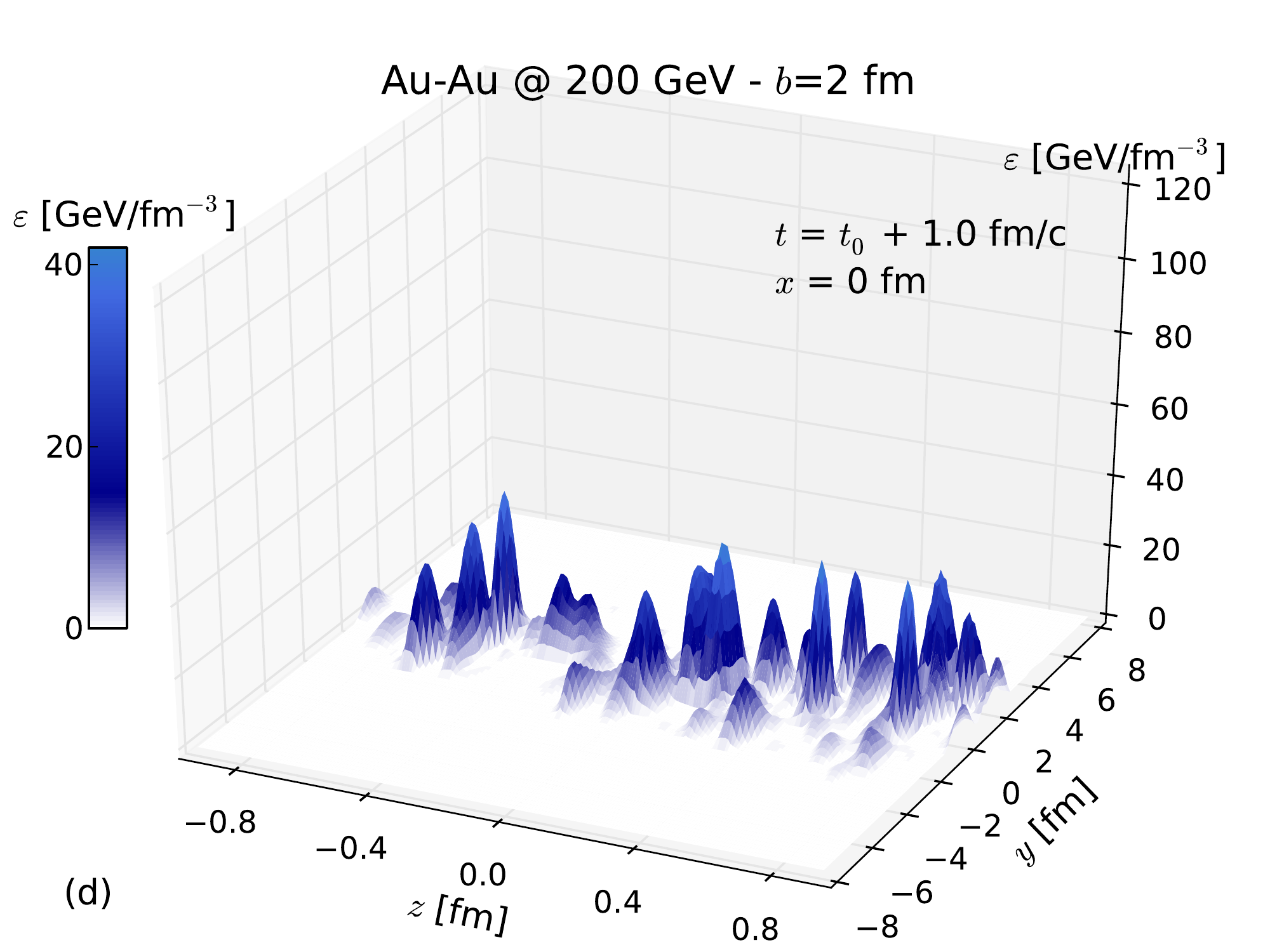}
  \end{center}
  \vskip -4mm
  \caption{(Color online) Energy density from PHSD for Au--Au collisions at $\sqrt{s_{NN}} = 200$ $A$GeV and $b = 2$ fm ($\sim$ 0\%--5\% centrality) in the transverse plane (a),(c) and the longitudinal plane (b),(d) and at the QGP formation time ($t \approx t_0 + 0.5$ fm/$c$) (a),(b), and a later time ($t = t_0 + 1.0$ fm/$c$) (c),(d). \label{Edens}}
  \vskip -3mm
\end{figure*}

\subsection{The parton-hadron string dynamics approach}
\vspace*{-2mm}

The PHSD approach \cite{Cassing2009,Bratkovskaya2011a} is a microscopic covariant transport model that incorporates effective partonic as well as hadronic degrees of freedom. The transition between partonic and hadronic matter is described dynamically by the formation of hadronic resonances. While the hadronic part is essentially equivalent to the conventional hadron-string-dynamics (HSD) approach \cite{Ehehalt1995,Cassing1999} the partonic dynamics is based on the DQPM \cite{Peshier2004,Cassing2007}, which describes QCD properties in terms of single-particle Green's functions (in the sense of a two-particle irreducible two-particle interaction (2PI) approach). For an overview of the thermodynamical properties (equation of state and transport coefficients) of the DQPM, see Ref. \cite{Marty2013}.

After initializing the nucleons by a Wood-Saxon distribution, for the binary collisions between projectile and target nucleons the {\sc FRITIOF} \cite{Andersson1998} and {\sc PYTHIA} \cite{Sjostrand2006}, respectively are employed. These form strings that are represented by leading hadrons and prehadrons. The leading hadrons can collide with further nucleons, however, with a reduced cross section (1/3$\sigma_{NN}$ for leading mesons and 2/3$\sigma_{NN}$ for leading baryons). In these collisions new strings are formed, which decay as well using the {\sc FRITIOF} model.

The energy, which the leading hadrons have lost, is converted into strings which are dissolved according to the DQPM parton spectral functions for local energy densities above $\epsilon_c$ ($\approx$ 0.5 GeV/fm$^3$). These partons are heavy: hence, their formation time is short and the QGP is formed shortly after the nuclei have passed through each other (at top energies currently available at RHIC). For convenience, we extract the initial condition at $t = t_0 + 0.5$ fm/$c$ --$t_0$ being the time of the first hard $NN$ collision-- for $\sqrt{s_{NN}}$ = 200 GeV. At this point, all strings are already melted and partons are formed which start to interact strongly.

\vspace*{-4mm}
\subsection{Initial conditions}
\vspace*{-2mm}

\begin{figure*}
  \begin{center}
    \includegraphics[width=8.5cm]{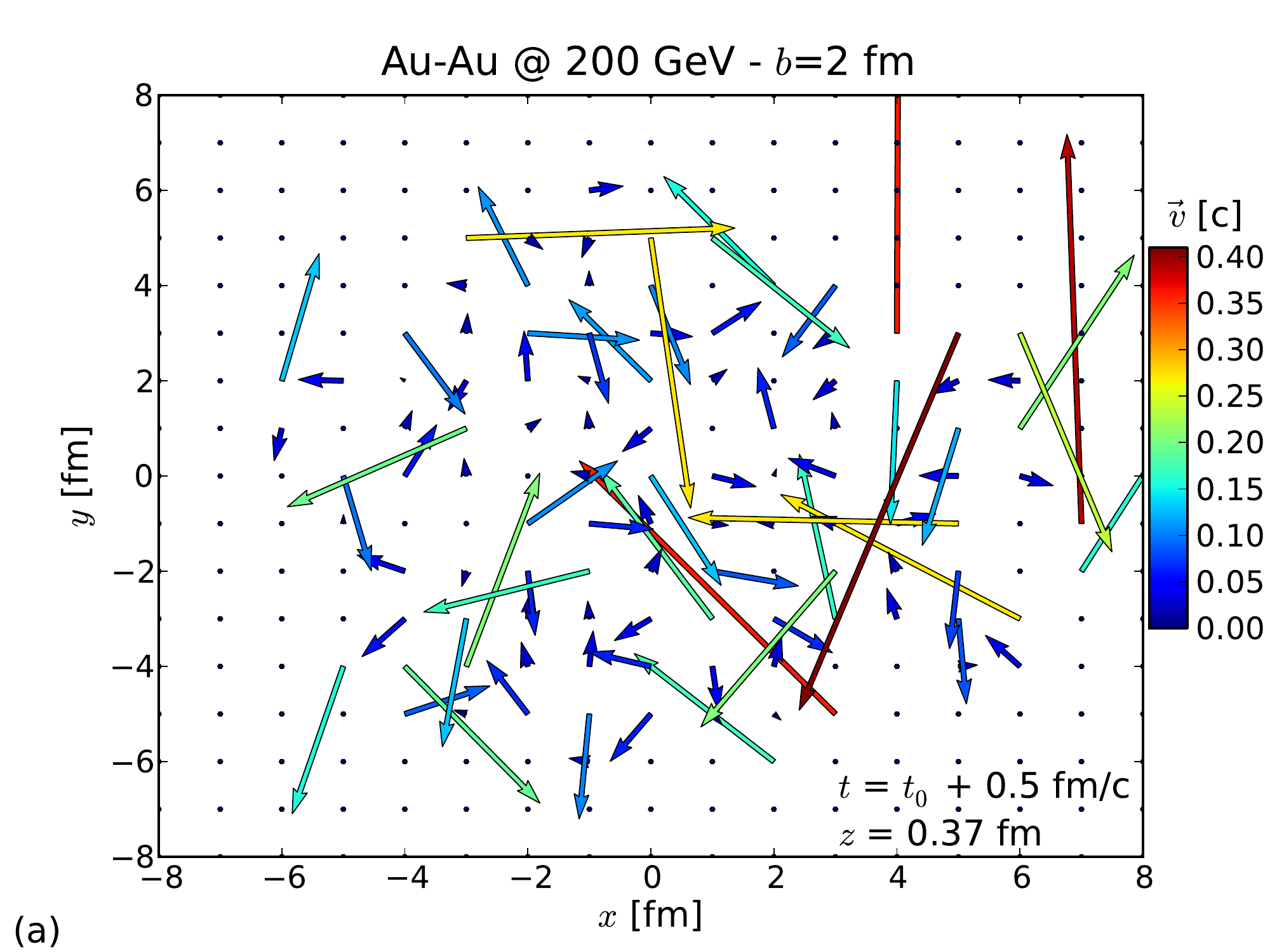} \ \ \ \
    \includegraphics[width=8.5cm]{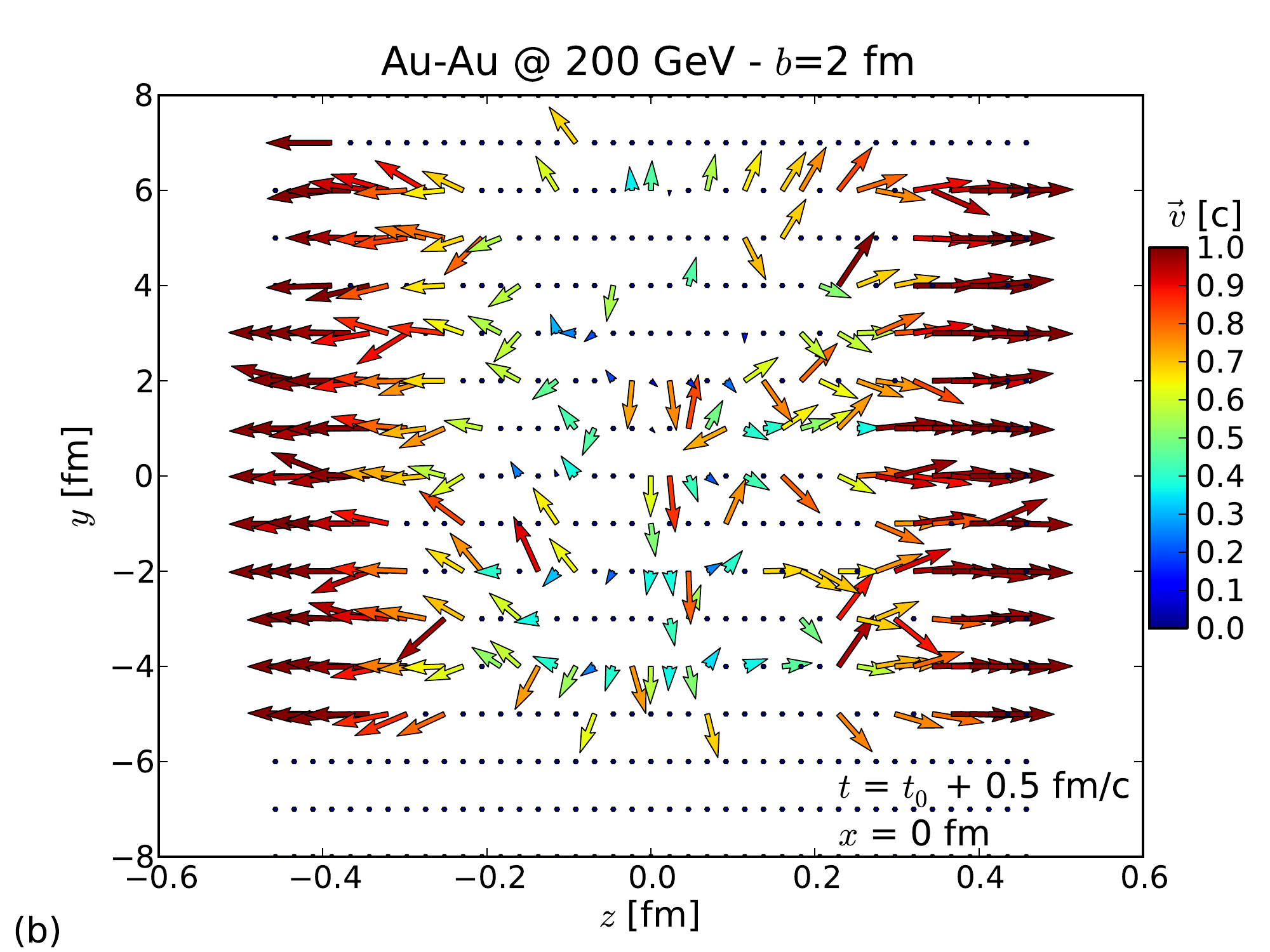}
  \end{center}
  \vskip -3mm
  \caption{(Color online) Cell velocities in PHSD for a Au--Au collision at $\sqrt{s_{NN}} = 200$ GeV and $b = 2$ fm (0\%$-$5\% centrality), in the transverse direction ($z = 0.37$ fm) (a) and in the longitudinal direction ($x = 0$ fm) (b).\label{Velocity}}
  \vskip -1mm
\end{figure*}

\begin{figure*}
  \begin{center}
    \includegraphics[width=8.5cm]{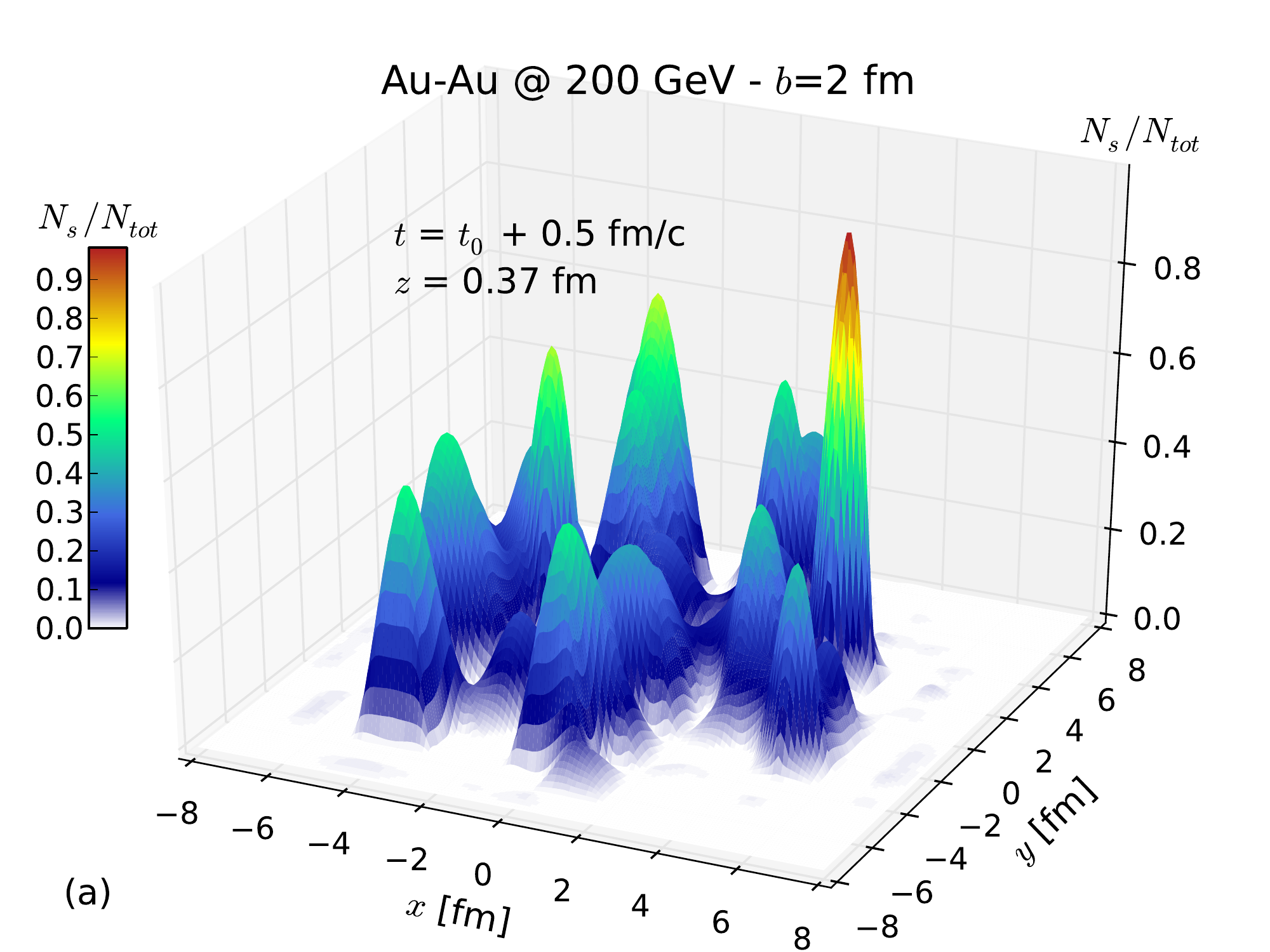} \ \ \ \
    \includegraphics[width=8.5cm]{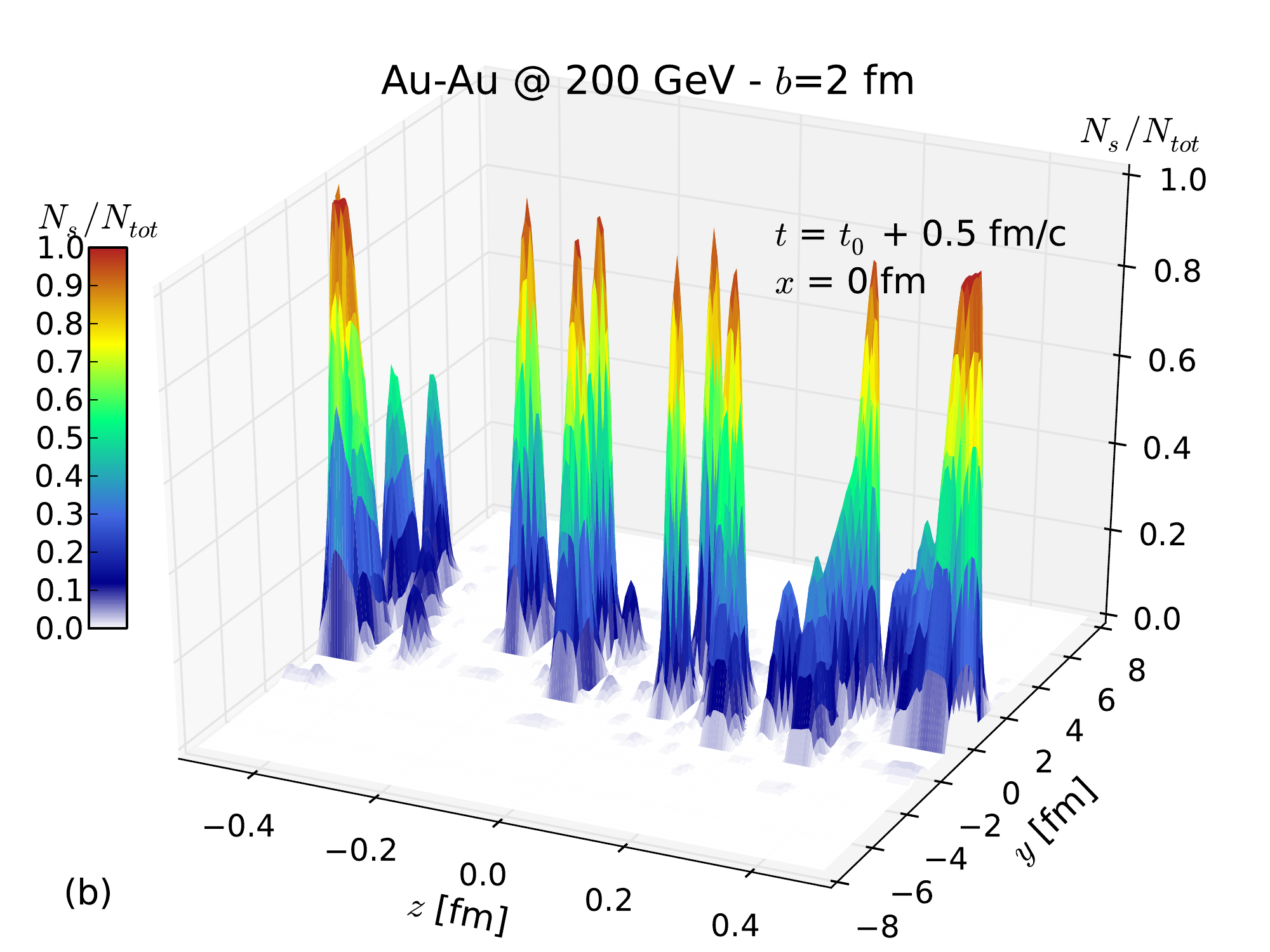}
  \end{center}
  \vskip -3mm
  \caption{(Color online) Strangeness abundance from PHSD for a Au--Au collision at $\sqrt{s_{NN}} = 200$ GeV and $b = 2$ fm (0\%$-$5\% centrality) in the transverse plane (for $z = 0.37$ fm) (a) and in the $z$-$y$ plane  (for $x = 0$ fm) (b).\label{Strangeness}}
  \vskip -1mm
\end{figure*}

At the time $\tau_0 + 0.5$ fm/$c$, when the nuclei have penetrated each other (owing to their Lorentz contraction) and strings are melted into DQPM partons, we synchronize both models. At that time the creation of the heavy DQPM partons gives large fluctuations in the particle distributions in a single event. We illustrate these fluctuations by displaying in Figs. \ref{Edens}$-$\ref{Strangeness} the energy density, the velocity, and the strangeness profiles, respectively, for a single Au-Au event at $\sqrt{s_{NN}} = 200$ GeV and $b = 2.2$ fm (0\%$-$5\% centrality).

At that time we calculate the energy-momentum tensor,
\begin{equation}
  T^{\mu\nu} = g \int\limits_{0}^{\infty} \frac{d^3 p}{(2\pi)^3} f(E) \frac{p^\mu p^\nu}{E},
\end{equation}
\vspace*{-3mm}
\begin{equation}
  T^{\mu\nu} =
  \begin{pmatrix}
    \varepsilon & Q_x & Q_y & Q_z \\
    -Q_x & P_x & \pi^{xy} & \pi^{xz} \\
    -Q_y & -\pi^{xy} & P_y & \pi^{yz} \\
    -Q_z & -\pi^{xz} & -\pi^{yz} & P_z
  \end{pmatrix},
\end{equation}
with the energy density $\varepsilon$, the pressure $P$, the momentum density $Q$, and the shear stress $\pi$. We notice that the energy-momentum tensor obtained in PHSD is not diagonal and therefore the particles do not represent an ideal fluid. The fluid is viscous and $\pi \ne 0$. We refer to  Refs. \cite{Ozvenchuk2013,Marty2013} for a further discussion of viscosity in the DQPM. We also compute the particle density in the local rest frame,
\begin{equation}
  n = g \int\limits_{0}^{\infty} \frac{d^3 p}{(2\pi)^3} f(E),
\end{equation}
which allows for extracting the baryon density and the flavor decomposition.
\vspace*{6mm}

The energy density of each cell in PHSD in the global calculation frame is given by
\begin{equation}
  \varepsilon = \frac{\sum_i E_i}{V_{\text{cell}}},
\end{equation}
with $\sum_i$ standing for the sum over particles in the current cell and $V_{\text{cell}} = \Delta x \Delta y \Delta z$ the volume of the cell. We note that in PHSD, $\Delta x = \Delta y = 1$ fm, and $\Delta z = 1 / \gamma_{cm}$ fm, $\gamma_{cm}$ being the Lorentz $\gamma$ factor for the transformation into the center-of-mass of the colliding nuclei. We can compute the energy density for each cell in the local rest frame using the velocity of this cell, which is
\begin{equation}
  \vec{\beta} = \frac{\sum_i \vec{p}_i}{\sum_i E_i}.
\end{equation}
This velocity gives the Lorentz factor $\gamma = 1/\sqrt{1 - \beta^2}$, and then the energy density of the cells in local rest frame is
\begin{equation}
  \varepsilon' = \frac{E'}{V'} = \frac{E / \gamma}{V \times \gamma} = \frac{\varepsilon}{\gamma^2}.
\end{equation}

On the left side of Figs. \ref{Edens}(a) and \ref{Edens}(c) we display the energy density in the transverse plane in the local rest frame of the cells, and on the right side we show the energy density in the longitudinal plane. The figures on the top [(a) and b)] are snapshots of the initial condition at a time $\tau_0 + 0.5$ fm/$c$ whereas the bottom figures display these quantities at $\tau_0 + 1.0$ fm/$c$. Thus, we see a fast decrease of the initial energy density, whereas the geometry of fluctuations in coordinate space is almost conserved. One can see that in any case the fluctuations are important and the granularity is in between $1/m_\pi \approx 1.5$ fm/$c$ and $1/m_q \approx 0.3$ fm/$c$.

In Figs. \ref{Velocity}(a) and \ref{Velocity}(b), we depict the longitudinal velocity $v_L$ and transverse velocity $v_T$ of the cells. The longitudinal velocity profile shows that the farther the cells are from the ``center'' ($z=0$ fm) the faster they are (the space-time rapidity is proportional to the rapidity). On the contrary, the azimuthal angle of transverse velocity is uniformly distributed. The absolute value of the transverse velocity is small compared to that in the longitudinal direction. This profile of the velocity distributions is obtained after the conversion of strings into pre-hadrons and the subsequent interactions of the DQPM partons respecting in each step the conservation of the 4-momentum as well as the flavor currents. This procedure assures as well that there is initially no transverse flow in the PHSD approach.

Figure \ref{Strangeness} shows how the flavor is distributed initially. We display the strangeness ratio, i.e., the number of strange and antistrange quarks over the total number of quarks in each cell $(N_s+N_{\bar s}) / N_{\text{tot}}$ in the transverse plane (a) as well as in the $z-y$ plane (b) (for $x$=0). We already know from experiment that at energies currently available at RHIC the chemical equilibration of strange quarks is not achieved, so the total number of strange quarks does not correspond to the thermal equilibrium value. In the $y-z$ direction (b) the mean value is --as expected-- close the center of the interaction, but the ratio fluctuates substantially. In longitudinal direction the strangeness is distributed over the whole interaction region (a).

We can summarize these observations as follows.
\begin{itemize}
  \setlength\itemsep{-3.5pt}
  \item[(i)] We have a strong energy deposition from leading participant baryons. In coordinate space this energy density fluctuates strongly.
  \item[(ii)] The longitudinal distribution of the partons is almost constant over the reaction region, reminiscent of the Bjorken scaling.
  \item[(iii)] The cell velocities are initially small; thus, a radial flow has to develop later.
   \item[(iv)] Neither globally nor locally is the energy-momentum tensor is diagonal. PHSD propagates the information of the non diagonal parts, in contradistinction to ideal hydrodynamical approaches, which neglect the off-diagonal elements of $T^{\mu \nu}$.
\end{itemize}
These initial conditions of PHSD show that at the synchronization time the system is not in a local thermal equilibrium.

\section{The Nambu-Jona-Lasinio model}

The NJL model is based on a Lagrangian which respects the same symmetries than QCD. For three flavors it reads \cite{Klevansky1992}
\begin{equation}
  \begin{aligned}
    \mathscr{L}_{NJL}&= \bar{\psi} \left( i \partial\!\!\!\!/- m_0 \right) \psi\\
             &+ \ G \sum_{a=0}^8 \left[   \left( \bar{\psi}            \lambda^a \psi \right)^2
                                         + \left( \bar{\psi} i \gamma_5 \lambda^a \psi \right)^2 \right] \\
             &+ \ K \left[   \mathrm{det} \bar{\psi} \left( 1 - \gamma_5 \right) \psi
                              + \mathrm{det} \bar{\psi} \left( 1 + \gamma_5 \right) \psi \right].
  \end{aligned}
  \label{NJL_lagrangian}
\end{equation}
The free parameters are the bare masses of quarks for $2+1$ flavors $m_{0u} = m_{0d} \ne m_{0s}$, and the coupling constants $G$ for scalar/pseudoscalar mesons (determined by the pion mass in vacuum), and $K$ for the flavor mixing (determined by the masses of the mixed states $\eta$ and $\eta'$). The table of parameters that we use can be found in Ref. \cite{Marty2012}.

The NJL model is an approach for quark and antiquark degrees of freedom. The mass of gluons is assumed to be large as compared to the momentum transfer in interactions among quarks and therefore the interaction is reduced to a 4-point interaction with an effective coupling constant. The masses of quarks are calculated in mean-field approximation. Meson masses are the pole masses of the summation of $q/\bar{q}$ polarization loops. This approach allows for describing parton and meson properties as well as cross sections with the help of very few parameters \cite{Marty2012}, which can be fixed by particle properties in vacuum. All masses and cross sections from this model can be calculated for a finite temperature $T$ and at finite chemical potential $\mu$.

\begin{figure} [t]
  \begin{center}
    \includegraphics[width=8.5cm]{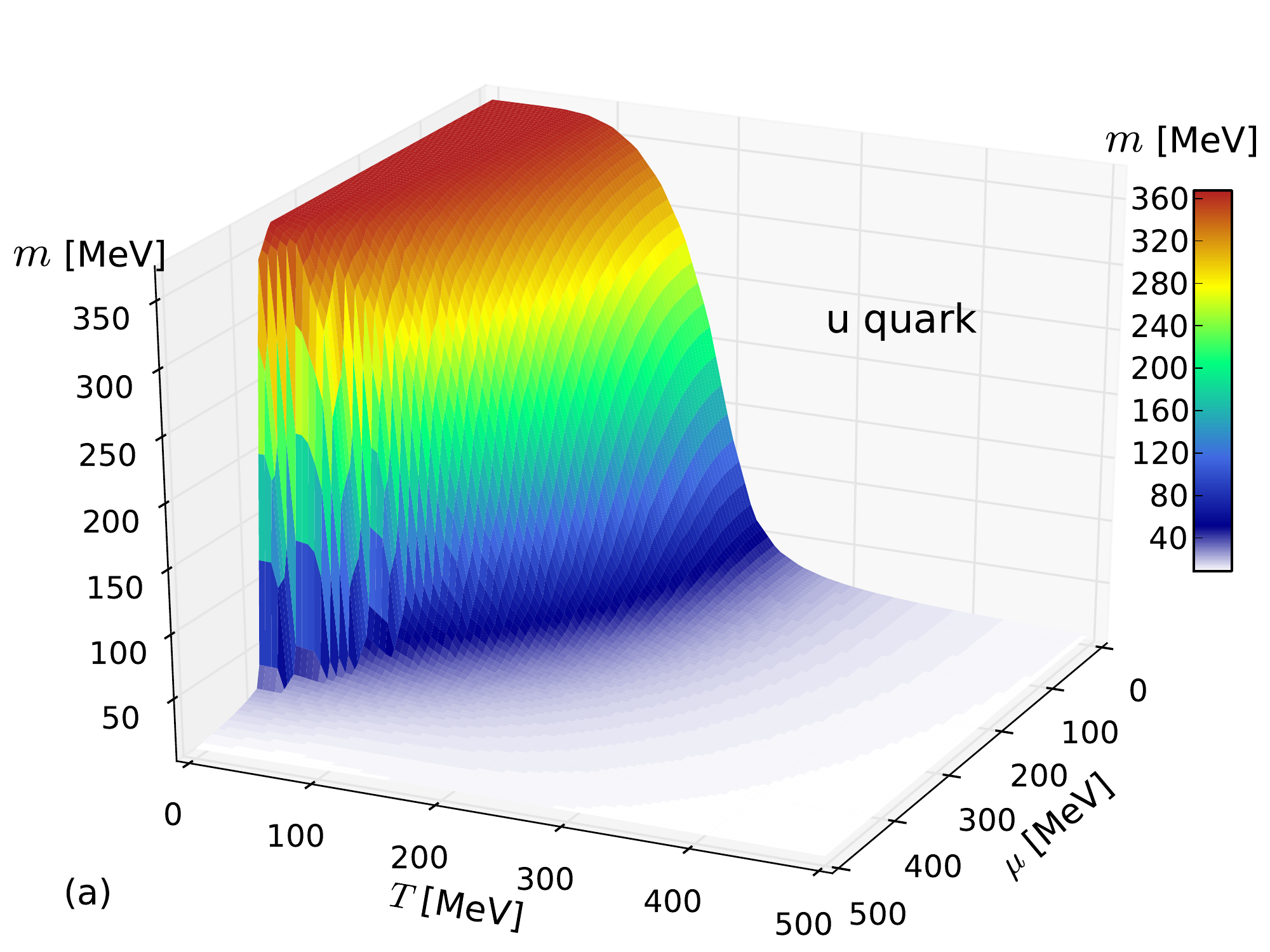}\\
    \includegraphics[width=8.5cm]{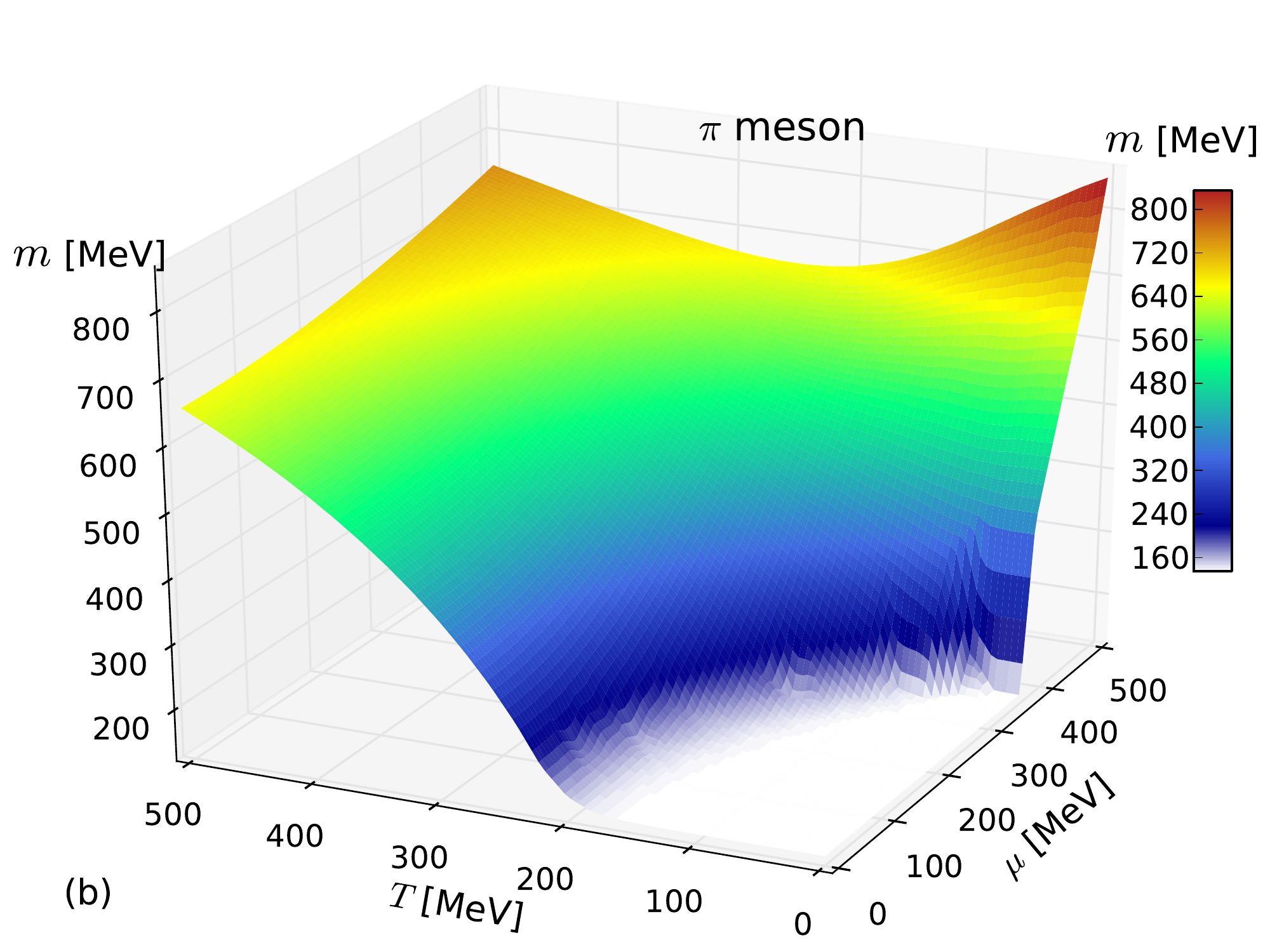}
  \end{center}
  \vskip -4mm
  \caption{(Color online) Masses of $u$ quark (a) and $\pi$ meson (b) as a function of $(T,\mu)$ in the NJL model. Note the different orientation of the axis in the upper (a) and lower (b) plots. \label{mass3D}}
  \vskip -2mm
\end{figure}

The masses of the light $u$ quark and of the $\pi$ meson are shown in Fig. \ref{mass3D} as a function of the temperature and the chemical potential. For small temperatures and chemical potentials, the mass of the quarks is dressed owing to the interaction with the scalar condensates, whereas the mass of the pion (being the Goldstone boson of the model) tends to its vacuum limit. When the temperature and/or the chemical potential increase(s), the scalar condensates disappear and thus the quark recovers its small bare mass, whereas the pion increases its mass and develops a finite width \cite{Marty2012}.

The hadronization cross sections as well as the elastic cross sections are strongly momentum-dependent and become very large owing to a $s$-channel resonance close to the Mott temperature, where $m_\pi = m_q + m_{\bar q}$. Consequently, the viscosity over the entropy density becomes small, $\eta/s \sim 0.1$ \cite{Marty2013}. Thus, during the expansion, close to $T_c$,  hadronization becomes important and leads finally to a gas of hadrons. The interaction among quarks is attractive, which reduces the slope of the transverse spectra during the expansion.

\begin{figure} [t]
  \begin{center}
    \includegraphics[width=8.5cm]{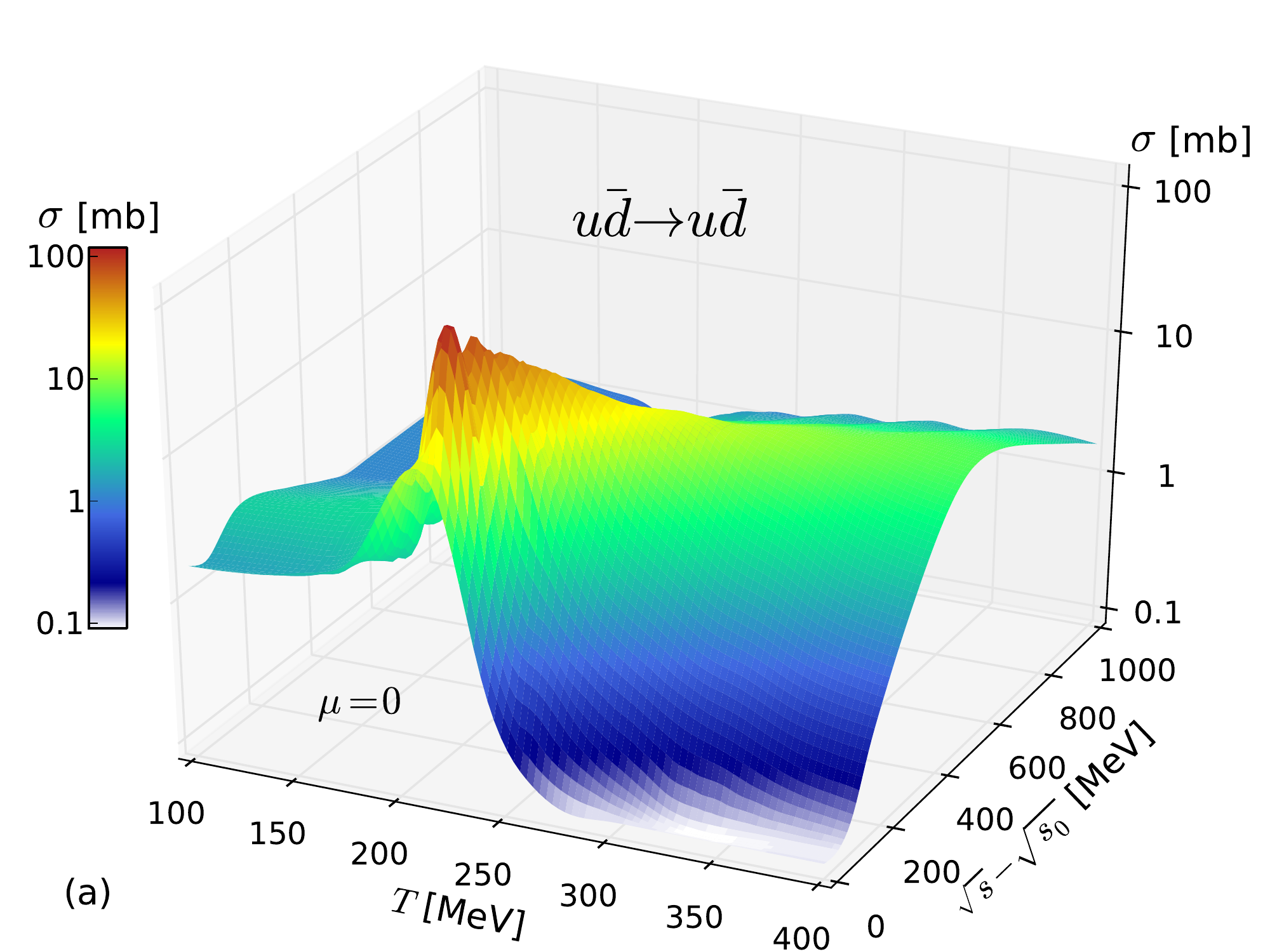}\\
    \includegraphics[width=8.5cm]{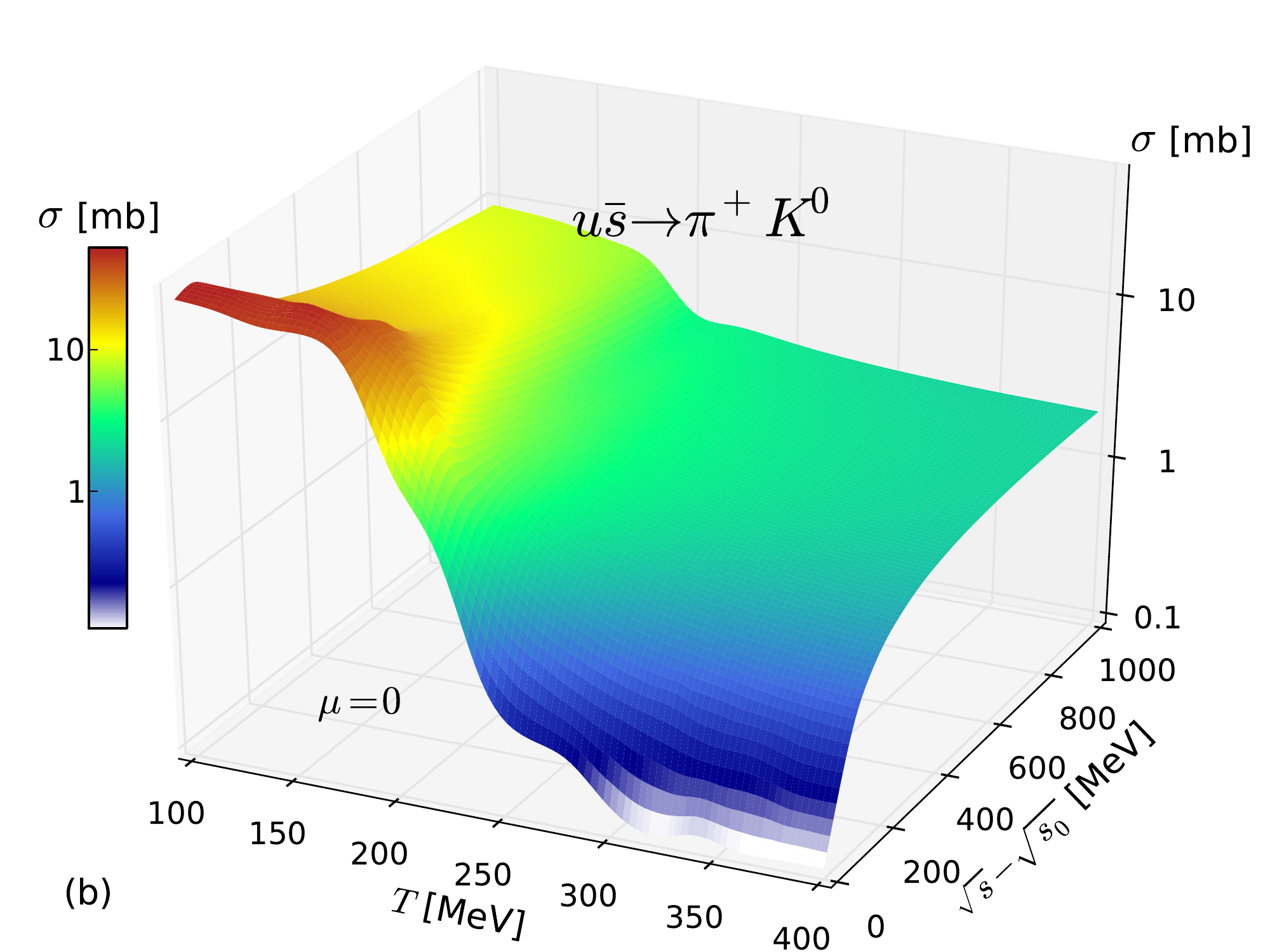}
  \end{center}
  \vskip -4mm
  \caption{(Color online) NJL cross sections for $u\bar{d} \to u\bar{d}$ (a) and $u\bar{s} \to \pi^+ K^0$ (b) as a function of $(T, \sqrt{s} - \sqrt{s}_0)$ in the NJL model, where $\sqrt{s}_0$ denotes the threshold center-of-mass energy.\label{CrossSection3D}}
  \vskip -3mm
\end{figure}

Figure \ref{CrossSection3D} shows the elastic cross section $q\bar{q}\to q\bar{q}$ and the hadronization cross section $q\bar{q}\to MM$ as a function of the temperature and the center-of-mass energy above threshold, $\sqrt{s_0}$. One clearly sees the peak of the cross section close to the critical temperature (around 200 MeV in this model) and close to the threshold for elastic scattering, which leads to a small $\eta/s$. The hadronization cross section shows that close to $T_c$ particles in the same phase-space region (having a small spatial distance) will hadronize. We discuss the hadronization method and its problems in detail in the Appendix.

\section{The transport code RSP}
\vskip -2mm

The transport code RSP is  based on the NJL Lagrangian, which we use for this investigation. This code is a relativistic $N$-body microscopic transport code (see Ref. \cite{Marty2012} for details). The relativistic dynamics model in RSP is referenced as INTEGRAL in Ref. \cite{Marty2012}.

This transport approach follows the ideas developed originally in Ref. \cite{Sorge1989} but differs in the constraints one imposes to reduce the $8N$-dimensional phase space to the $6N+1$-dimensional phase space in which particle trajectories can be defined. These constraints allow not only for describing frame-invariant equations of motions but also allow for a numerical solution of the equation of motion for an interacting $N$-body system as has been demonstrated in Ref. \cite{Marty2012}, where the reader may find all relevant details. The equations of motion in this approach are given by
\begin{equation}
  \frac{d q_i^\mu}{d \tau} = \frac{p_i^\mu}{E_i},
  \qquad
  \frac{d p_i^\mu}{d \tau} = - \sum_{k=1}^N \frac{1}{2 E_k}\frac{\partial V_k}{\partial {q_i}_\mu},
  \label{final_relativistic_equation_motion}
\end{equation}
which ensure to a causal dynamics which conserves total energy. $\tau$ is the clock time for the calculation to which all individual times of the particles are connected by constraints. The NJL model enters into these equations through the effective mass discussed above, coming from chiral symmetry breaking. Assuming a local equilibrium, the derivative of the potential is proportional to the derivative of the mass with respect to the local temperature and the spatial derivative of the temperature
\begin{equation}
\frac{1}{2 E_k} \frac{\partial V_k}{\partial {q_i}_\mu}=
\frac{m_k}{E_k} \frac{\partial m_k}{\partial {q_i}_\mu}=
\frac{1}{\gamma_k} \frac{\partial m_k}{\partial T_k}\frac{\partial T_k}{\partial {q_i}_\mu},
\end{equation}
where the temperature is given by the local particle density (see Ref. \cite{Marty2012} for more details).

This new transport code is designed to describe the plasma phase as well as the hadronization and the hadronic gas. Only quarks and antiquarks as well as $SU(3)$  pseudoscalar mesons are presently included  with the corresponding cross sections. An extension including vector mesons and baryons is currently developed. It does not provide the initial condition, i.e., the mechanism by which the plasma is formed. This it shares with other approaches for the expanding plasma like (ideal or viscous) hydrodynamics.

The most advanced hydrodynamical approaches for 3+1D ideal \cite{Schenke2010} and viscous case \cite{Schenke2011} seem to indicate that this method describes correctly the experimental results for a broad variety of systems, from the  RHIC beam energy scan \cite{Karpenko2013} up to LHC results for Pb+Pb \cite{Luzum2009} and even p+Pb \cite{Bozek2012}.

Comparable to the PHSD approach, the RSP model can describe the time evolution of the plasma even if initially it does not come to an (almost) local equilibrium by solving the time-evolution equations of the partons and not that for the energy density like in hydrodynamics. Being a $n$-body approach, the RSP allows for the description of the time evolution of non equilibrium fluctuations and goes beyond the possibilities of the PHSD model, which solves the dynamics on the level of a 2 PI approach). This may be of importance for strongly coupled plasmas at finite baryon chemical potential, where the NJL Lagrangian predicts a first-order phase transition.



\vspace*{-4mm}
\section{Smooth initial conditions}
\vspace*{-2mm}

Before discussing the complex fluctuating initial conditions from relativistic heavy-ion collisions, we compare the two transport codes using smooth partonic initial conditions, i.e., a fireball of quarks and antiquarks with some spatial eccentricity $\varepsilon$. In momentum space we employ a thermal distribution with temperature $T = 2 T_c$ and in coordinate space a smooth Gaussian distribution with a spatial eccentricity $\varepsilon$ resembling the shape of the fireball in nucleus-nucleus collisions at finite impact parameter. The initial number of partons roughly corresponds to RHIC events for Au-Au collisions at $\sqrt{s_{NN}}=200$ GeV. This set up allows us to study how the elliptic flow develops as a response to the initial eccentricity.

\begin{figure} [t]
  \begin{center}
    \includegraphics[width=7cm]{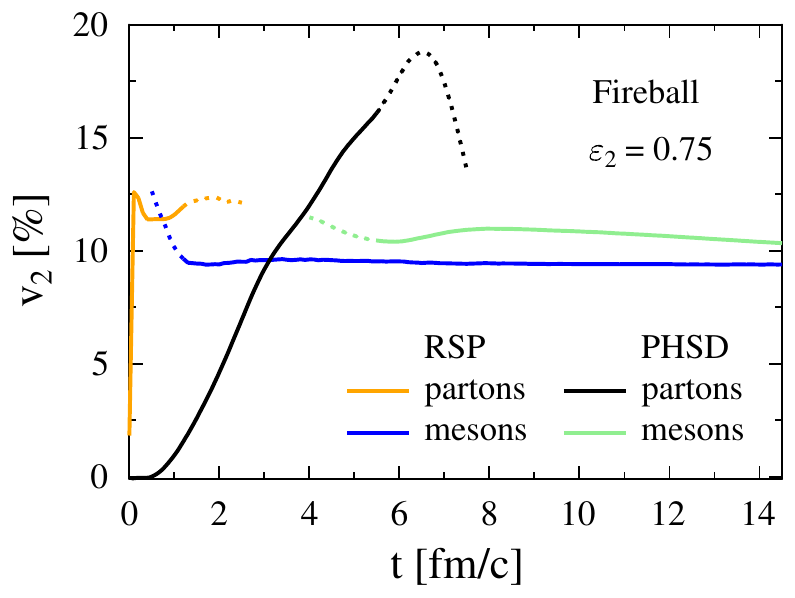}
  \end{center}
  \vskip -5mm
  \caption{(Color online) Time evolution of the elliptic flow $v_2$ for the expanding fireball in PHSD and RSP for smooth Gaussian initial conditions with a spatial eccentricity of $\varepsilon = 0.75$.\label{example1}}
  \vskip -3mm
\end{figure}

The time evolution of the elliptic flow $v_2$ is displayed in Fig. \ref{example1}. The $v_2$ is built up quite slowly, on the order of a couple of fm/$c$ ($10^{-23}$ s), in the PHSD approach in which partons can only interact after a formation time given by the inverse transverse mass in the rest frame of the particle. The evolution of $v_2$ is much faster in the RSP, where the particles start to interact as soon as the system starts to expand. In both approaches the $v_2$ of  partons is higher than that of the finally observed mesons. This is a kinematical effect: The mass of partons at the transition is higher than the mass of mesons; thus, the momentum increases at the phase transition and the flow decreases \cite{Cassing2008}. In PHSD the number of mesons is higher than the initial number of partons; this decreases additionally the flow and creates entropy.

\begin{figure} [b]
  \begin{center}
    \includegraphics[width=7cm]{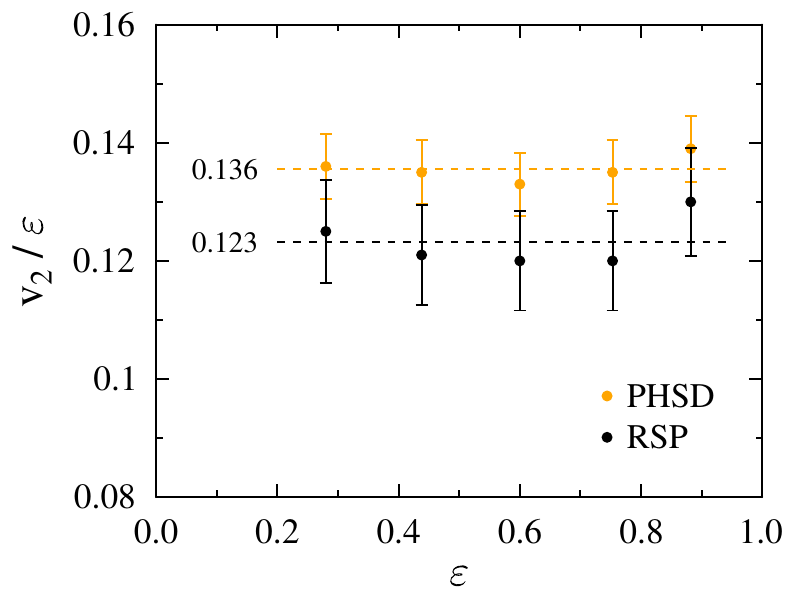}
  \end{center}
  \vskip -5mm
  \caption{(Color online) The elliptic flow over eccentricity ratio $v_2/\varepsilon$ as a function of the eccentricity, for the expanding fireball in PHSD and RSP for smooth Gaussian initial conditions.\label{example2}}
\end{figure}

Despite of the quite different time evolution and despite of the fact that the  DQPM and the NJL models show a different temperature dependence of the shear viscosity \cite{Marty2013}, the final flow $v_2$ of both approaches differs only by 10\% and the ratio $v_2/\varepsilon$ is almost independent of the eccentricity, as shown in Fig. \ref{example2}. A similar behavior has been observed in models that describe the plasma expansion by ideal hydrodynamical equations in Refs. \cite{Kolb2000,Alt2003}. Thus, both approaches, despite the very different properties of their constituents, generate some kind of hydrodynamical behavior under the condition of an expanding plasma and it will be difficult to conclude from the experimental results on the properties of the partons during the plasma expansion.

\begin{figure} [b]
  \begin{center}
    \includegraphics[width=7cm]{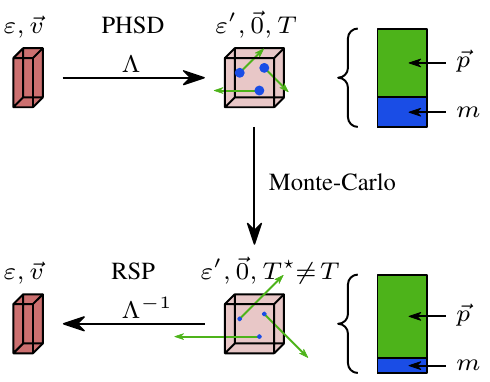}
  \end{center}
  \vskip -5mm
  \caption{(Color online) Conversion of energy density in the cell --where equilibrium is assumed-- from one model to another, knowing their equations of state and the real particle density.\label{conversion}}
\end{figure}

\section{Conversion of the PHSD initial condition}
\vspace*{-3mm}

\begin{figure*}
  \begin{center}
    \includegraphics[width=8.5cm]{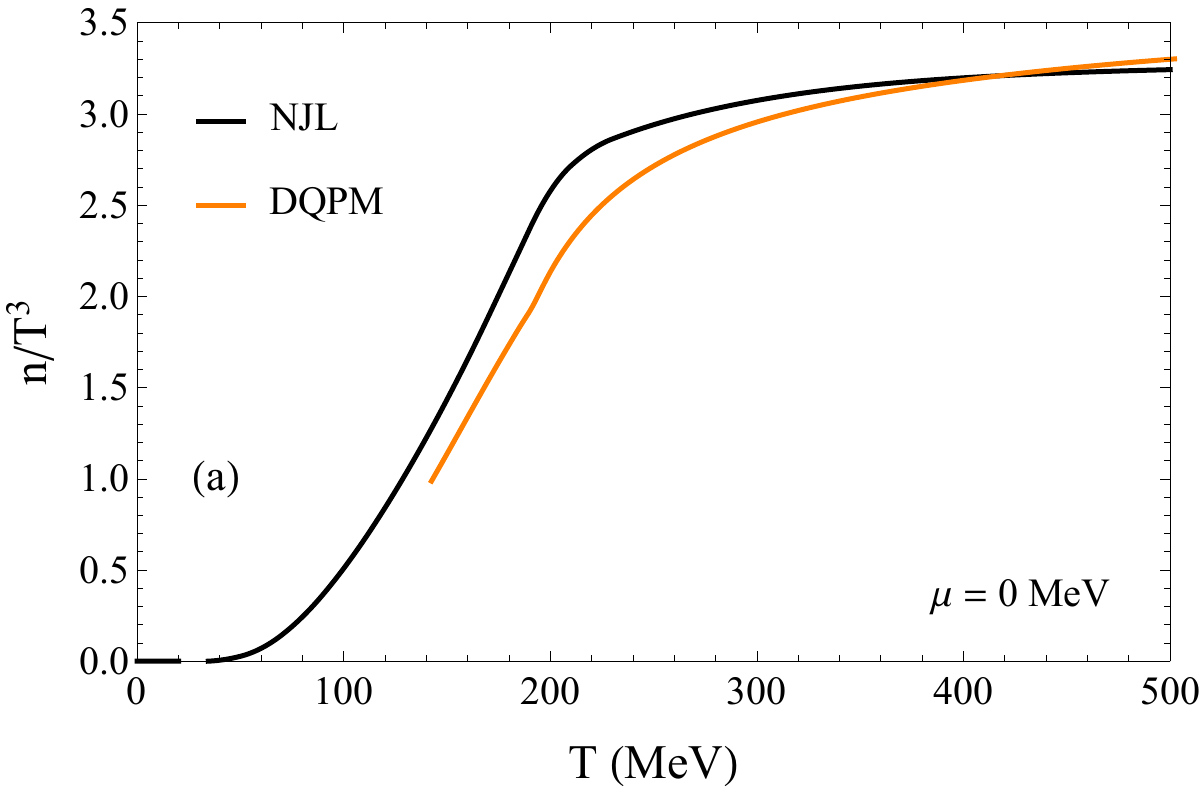} \ \ \
    \includegraphics[width=8.5cm]{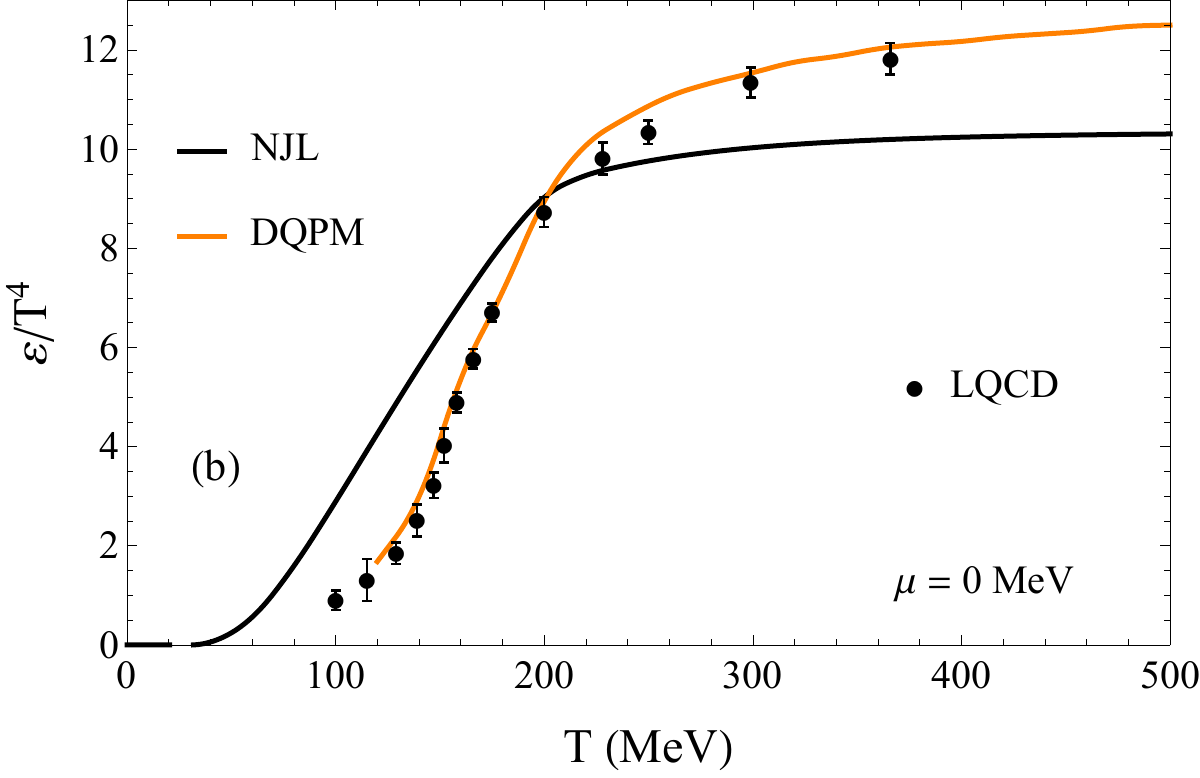}
  \end{center}
  \vskip -4mm
  \caption{(Color online) Equation-of-state comparison between NJL and DQPM, for the particle density (a) and energy density (b). Lattice QCD data are taken from Ref. \cite{Borsanyi2010}.\label{EoS_comp}}
  \vskip -3mm
\end{figure*}

The description of the time evolution of the QGP and its subsequent hadronization is one of the big challenges for the present-day transport approaches. One may assume that initially the energy-momentum tensor is diagonal, like it is assumed in ideal hydrodynamics or that the off-diagonal matrix elements are small, as assumed in viscous hydrodynamics. We want to compare whether very different descriptions for the time evolution of the system, the PHSD and the RSP approach, lead to a different time evolution of the system and finally to different distributions of the observables.

This is only possible if one starts in both approaches from the same initial conditions. Here we start out from an initial condition what is obtained by converting the Lund string model directly into partons described by the DQPM, an initial condition what is generated in the PHSD approach and what does not have the limitations of the hydrodynamical models. To use this PHSD initial condition for the RSP approach, we have to transfer it into the RSP model, a procedure that is not unique. We discuss two possibilities in this section.

\vspace*{-5mm}
\subsection{Equilibrium conversion}
\vspace*{-3mm}

The PHSD initial profile provides several ingredients which we can use for generating the NJL plasma: the energy density $\varepsilon$, the flow velocity ($\beta_x$, $\beta_y$, $\beta_z$), and the flavor abundance ($N_u$, $N_{\bar{u}}$, $N_d$, $N_{\bar{d}}$, $N_s$, $N_{\bar{s}}$), for each cell. Based on different physical assumptions, the conversion to microscopic degrees of freedom can be carried out in different ways. For the first conversion method we assume a local thermal equilibrium in the cell. This means that the momentum of the quarks is isotropically distributed in the cell's local rest frame according to a thermal (Fermi/Bose) distribution, and that the number of particles is determined by the energy density in the cell through the equation of state.

The procedure of conversion is depicted in Fig. \ref{conversion}. Starting from PHSD cells, we move them to the local rest frame and determine the local energy density. Using the equation of state of the NJL model (which is different from the DQPM equation of state \cite{Marty2013}) we can calculate from the local energy density $T$ and $\mu$ and subsequently the number of particles, the momentum distribution and the masses. Finally, we transform back the cells to the global calculational frame. The energy is conserved in this procedure, but the energy is differently distributed in the DQPM as compared to the NJL.

We only consider this procedure for cells in which the energy density in the local rest frame is above the critical limit of $\varepsilon_c \simeq 1$ GeV/fm$^3$, which corresponds to the NJL critical temperature of $T_c \simeq 200$ MeV \cite{Marty2013}. In the PHSD approach the other cells, presenting the corona of the collision, are not converted into partons. For the heavy-ion collisions --which we are considering in the rest of the paper--  this cut-off removes less than 5\% of the total energy of the system, which does not affect significantly the final results. For cells with a smaller energy density this method is not applicable because the fluctuations become too large.

As seen in Fig. \ref{conversion} the repartition in kinetic energy and potential energy (depicted, respectively, as $\vec p$ and $m$ on the right hand side of the figure) is not the same for both models, but also not far from each other. The interaction measure from the trace anomaly compared between the DQPM and NJL shows a difference \cite{Marty2013}, but the effect of this difference is exactly what we want to observe with the same initial energy-density profile.

In Fig. \ref{EoS_comp} we display the particle density $n/T^3$ and the energy density $\varepsilon/T^4$ as a function of temperature for both models. We voluntarily do not normalize by the Stefan-Boltzmann limit to emphasize the fact that these equations of state are not very different in the range $T_c < T < 2 T_c$ for the particle density. For the energy density, the DQPM equation of state has been fitted to recent lattice QCD data, while the NJL model has been adjusted to older lattice data and does not include gluon degrees of freedom, which explains the difference.

This conversion is noted in the figures as method 1.

\vspace*{-5mm}
\subsection{Out-of-equilibrium conversions}
\vspace*{-3mm}

The equilibrium conversion does not take into account several important properties of the PHSD initial condition like the particle density after string decay and the momentum distribution in the cells in the local rest frame. To study its influence we investigated a second conversion procedure which respects these properties.

\begin{figure} [b]
  \vspace*{-2mm}
  \begin{center}
    \includegraphics[width=8.5cm]{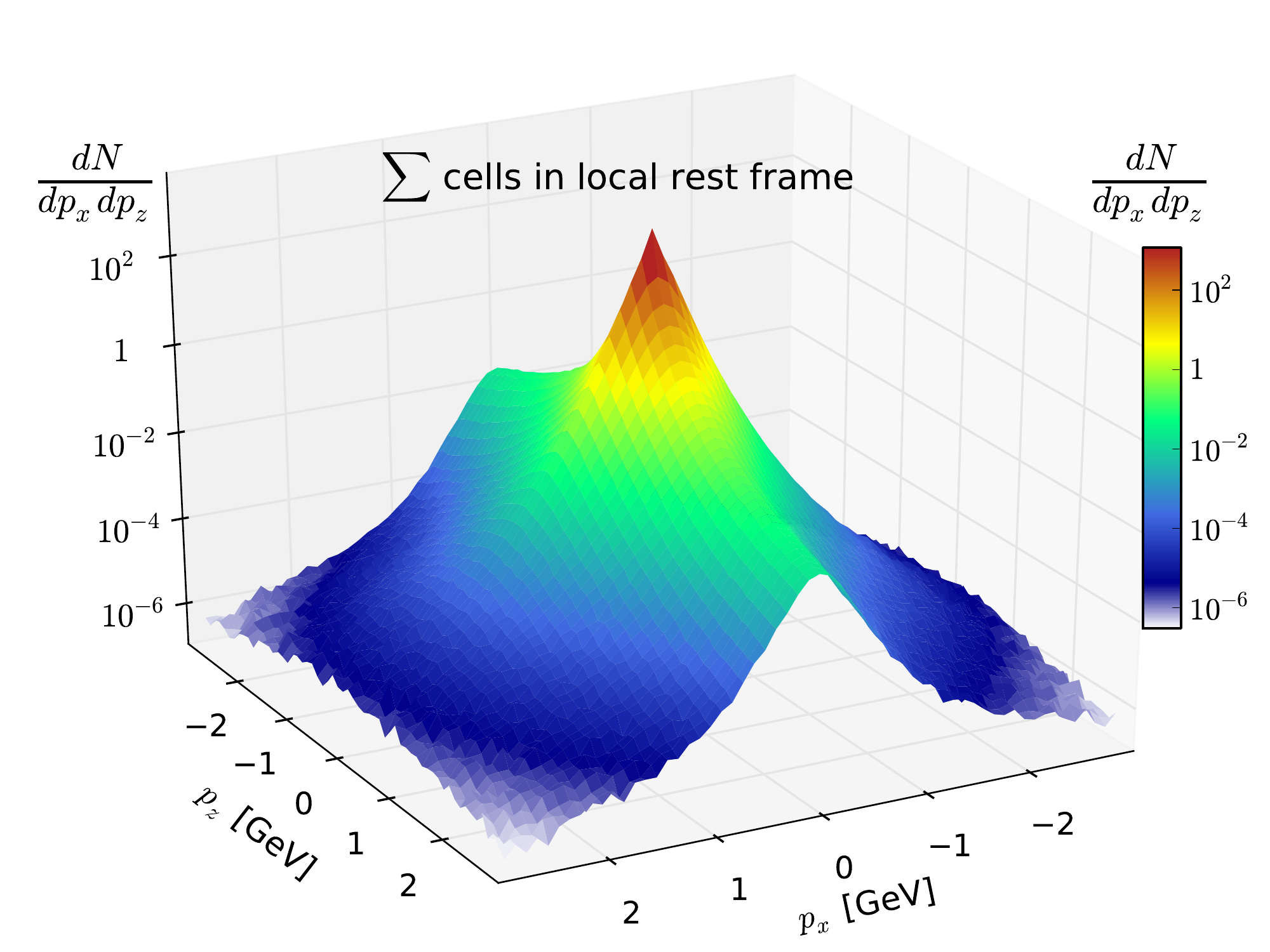}
  \end{center}
  \vskip -4mm
  \caption{(Color online) Probability distribution of particle momenta  ($p_z,p_x$) in a cell at rest.\label{multiplicity_ptpz}}
\end{figure}

In Fig. \ref{multiplicity_ptpz} we display the probability distribution to find a momentum in the $z$ and $x$ directions in the rest frame of the local cells of the PHSD approach. It is clearly seen that the distribution is not isotropic for large values of $\vec p$. Nevertheless, for the thermal moment ($\vec p_{th} < 1.5$ GeV) and hence for the majority of particles in the cells the distribution is almost isotropic, and follows a Boltzmann distribution. The mean value of the momentum is, however,  far below the one we expect for the temperature given by the energy density. If we take this steeper distribution of $p$ into account, we have to increase the local particle density to obtain the right energy density.

We start out from the energy density in the cells, which corresponds, using the DQPM equation of state, to a temperature $T$. Because the energy density is conserved, this energy density gives us, using the equation of state from the NJL in Fig. \ref{EoS_comp}, the temperature $T^\star$. With help of the DQPM equation of state we can also convert the particle density in the cells into a temperature and find $\hat{T} \ne T$. To calculate the particle density in NJL  we use the temperature
\begin{equation}
  \hat{T}^\star = \frac{\hat{T}}{T} T^\star.
\end{equation}
Thus, the total number of particle is changed compared to the equilibrium value such as
\begin{equation}
  n' = \int_0^\infty \alpha f_{eq}(p)\ d^3 p = \alpha n.
\end{equation}
The observable $\alpha$ indicates how far the particle density is from the equilibrium value. For the events we are considering we find that $\alpha \simeq 1.3$ -- 1.8. To conserve the total energy density the average energy per particle is changed accordingly,
\begin{equation}
  \varepsilon' = \int_0^\infty \alpha^{-1} p\ \alpha f_{eq}(p)\ d^3 p = \varepsilon.
\end{equation}

\begin{figure} [t]
  \begin{center}
    \includegraphics[width=7cm]{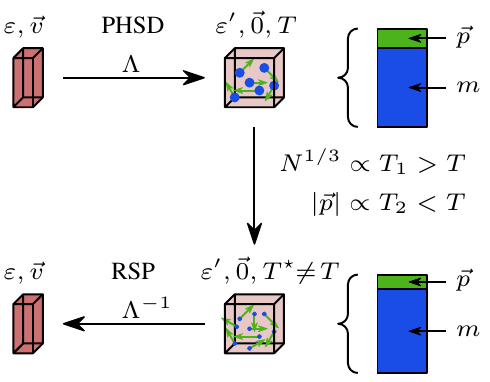}
  \end{center}
  \vskip -5mm
  \caption{(Color online) Out-of-equilibrium conversion of energy density in the cell from one model to another, knowing their equations of state and the real particle density.\label{conversion_new}}
  \vskip -3mm
\end{figure}

This out-of-equilibrium conversion is summarized in Fig. \ref{conversion_new}. The effective temperatures $T$ and $\hat T$ are calculated for each cell and they are used to determine the density and the momentum distribution of the RSP partons.

The consequence of such an out-of-equilibrium initialization is depicted on the right-hand side of Fig. \ref{conversion_new}. The potential part, given by the number of massive particles, is larger than previously, and, consequently, the effect of the mean-field interaction is expected to be much more important. We discuss this effect in the next section.

In this conversion, the number of NJL particles in the cells is larger than in equilibrium and is very close to that of PHSD. Indeed, the initial number of particles in PHSD is very close to the final number of hadrons (assuming that each quark/diquark convert into one hadron). With this out-of-equilibrium initialization, this is also the case for the NJL particles. This picture is different from the standard approach in hydrodynamical calculations, where the particle number is smoothly increasing during expansion. This conversion is noted in the figures as method 2.

Because in the above approach the initial particle number is very close in PHSD and in RSP we prepared a third method of conversion for initial conditions. In this method we directly convert DQPM partons into NJL quarks and antiquarks using the relation
\begin{equation}
  |p_1|^2 + m_1^2 = |p_2|^2 + m_2^2.
\end{equation}
Thus, we conserve the initial position/momentum correlations between the particles by  just shifting the absolute value of the momentum to accommodate the mass difference between NJL and DQPM parton masses.
PHSD gluons are converted into $\pi^0$'s, which decay in the plasma or collide to give $\pi^0 \pi^0 \to q \bar q$. This conversion is noted in the figures as method 3.

\begin{figure*}
  \begin{center}
    \includegraphics[width=8cm]{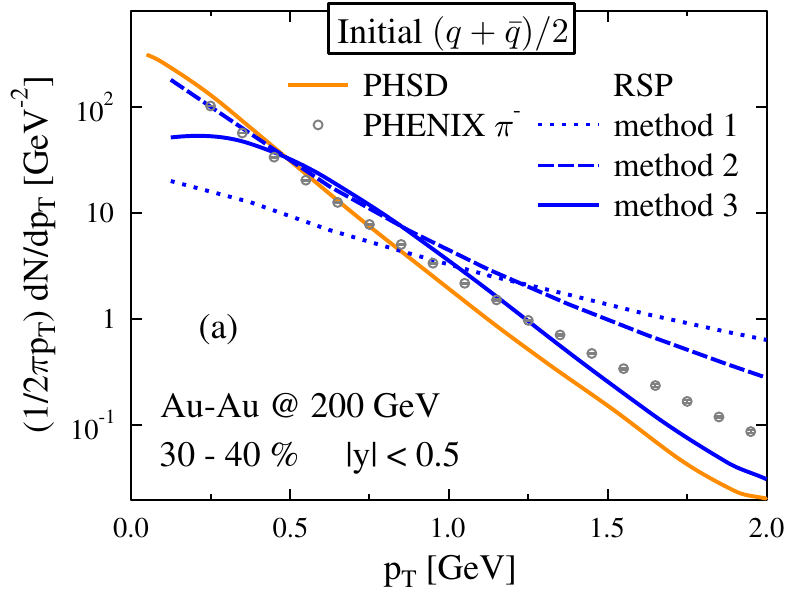} \ \ \ \
    \includegraphics[width=8cm]{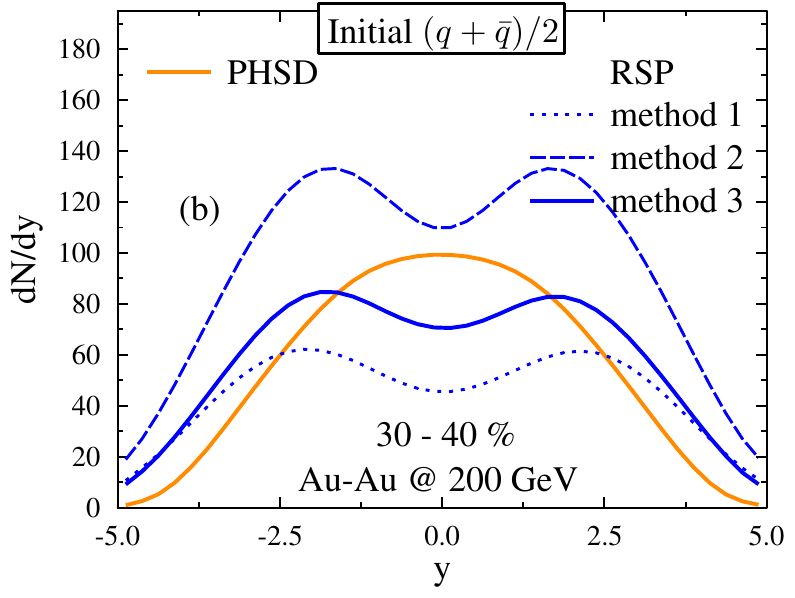}\\[2mm]
    \includegraphics[width=8cm]{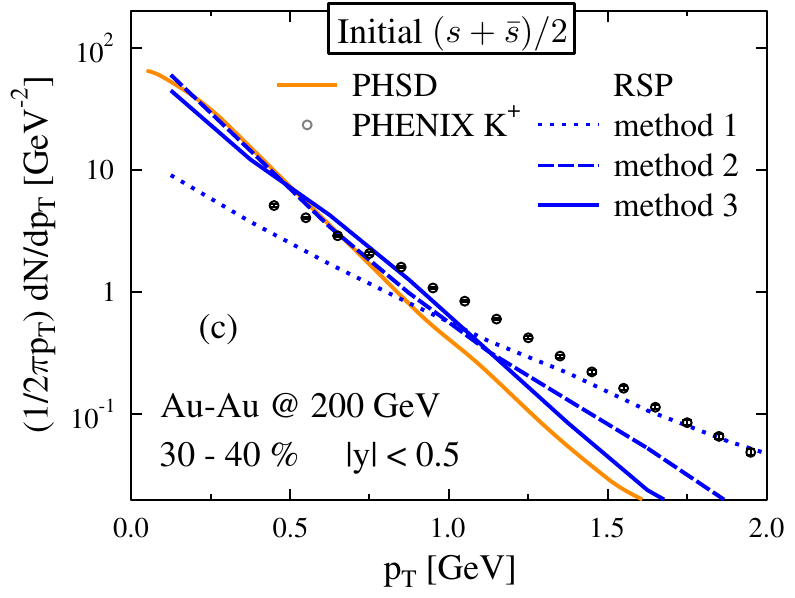} \ \ \ \
    \includegraphics[width=8cm]{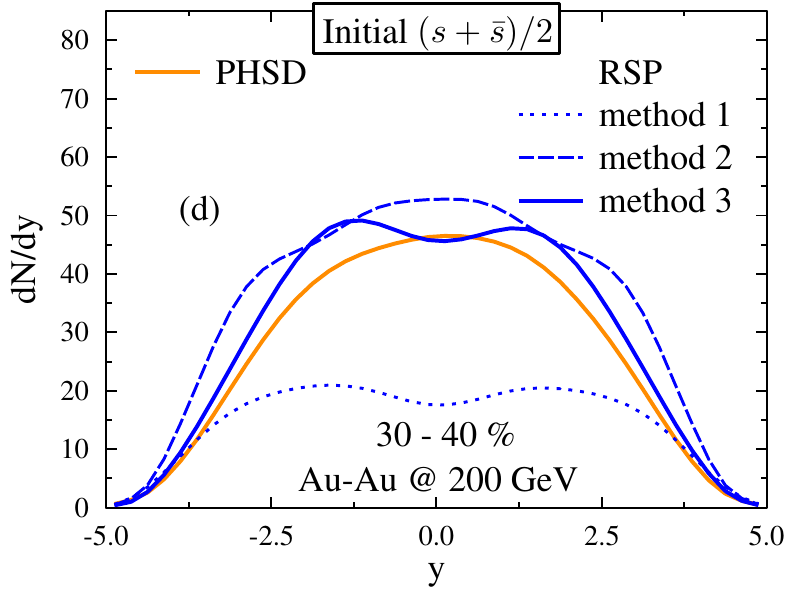}
  \end{center}
  \vskip -4.5mm
  \caption{(Color online) Initial distributions of the quarks in PHSD and RSP, using the three different conversion methods explained in the text. We display the transverse momentum spectrum for $|y| < 0.5$ and the rapidity distribution, for light quarks in the top panels [(a) and (b)] and for strange quarks in the bottom panels [(c) and (d)]. \label{results_initial1}}
  \vskip -4mm
\end{figure*}

\vspace*{-5mm}
\section{Final Results}
\vspace*{-3mm}

To see the influence of the different descriptions of the expansion we compare the transverse momentum and rapidity spectra of PHSD and RSP for the reaction Au-Au at $\sqrt{s_{NN}} = $200 GeV for 30\%$-$40\% centrality ($b = 8.4$ fm). We display the initial distribution of the quarks and compare them with the final distribution of the mesons. We display as well the elliptic flow, $v_2$, for the two approaches.



The results of RSP are presented for the three methods for the conversion of the initial profile of PHSD into a initial distribution of RSP which was explained in the last section:
\begin{itemize}
  \setlength\itemsep{-3.5pt}
  \item[(i)] the equilibrium conversion assuming local equilibrium in each cell in RSP (method 1);
  \item[(ii)] the out-of-equilibrium conversion taking into account that the particle density in PHSD is not thermal (method 2);
  \item[(iii)] the direct conversion in which each PHSD parton has the same energy as a parton in RSP (method 3).
\end{itemize}

Figure \ref{results_initial1} displays the initial spectra of quarks in PHSD and in RSP for the different conversion methods. Here $q,\bar q$ indicates the average over light quarks and antiquarks and $s,\bar s$ indicates the average over strange quarks and antiquarks. We display as well the PHENIX data for reference \cite{Phenix2004} to make the final and initial distribution easier to compare.

\begin{figure*}
  \begin{center}
    \includegraphics[width=8cm]{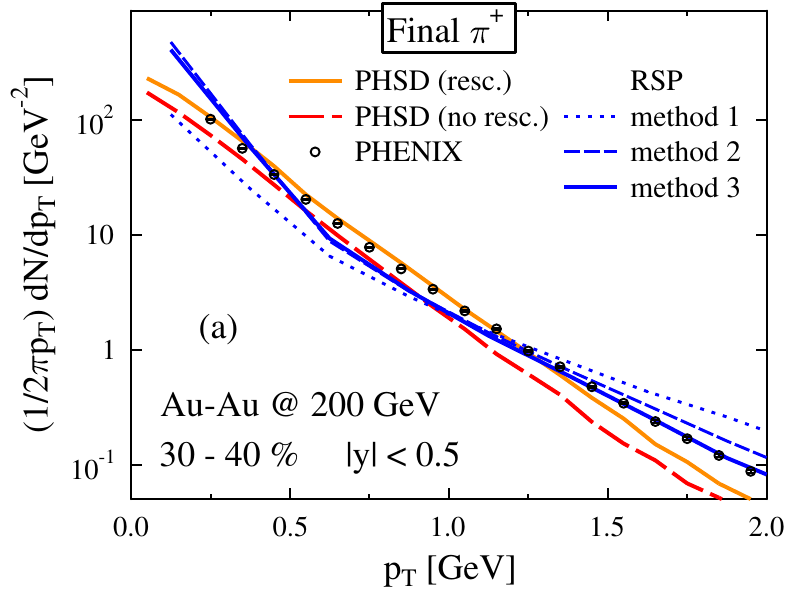} \ \ \ \
    \includegraphics[width=8cm]{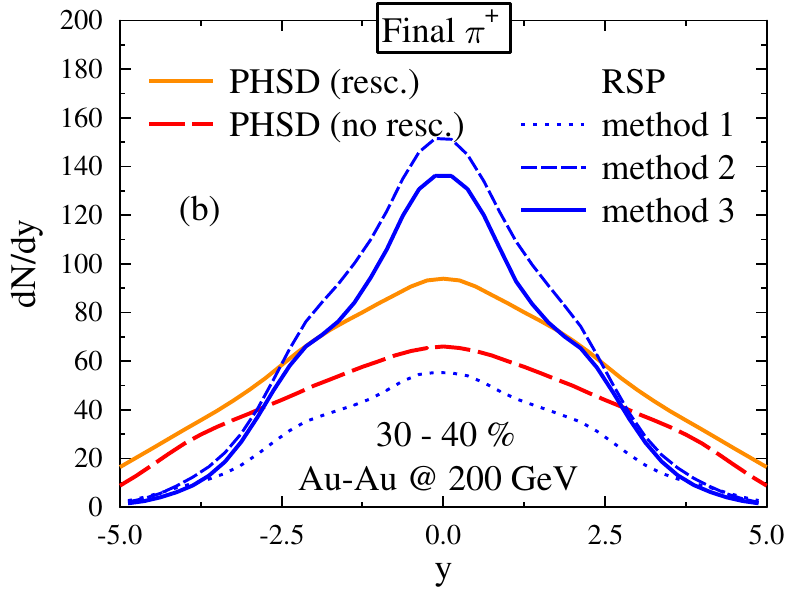}\\[2mm]
    \includegraphics[width=8cm]{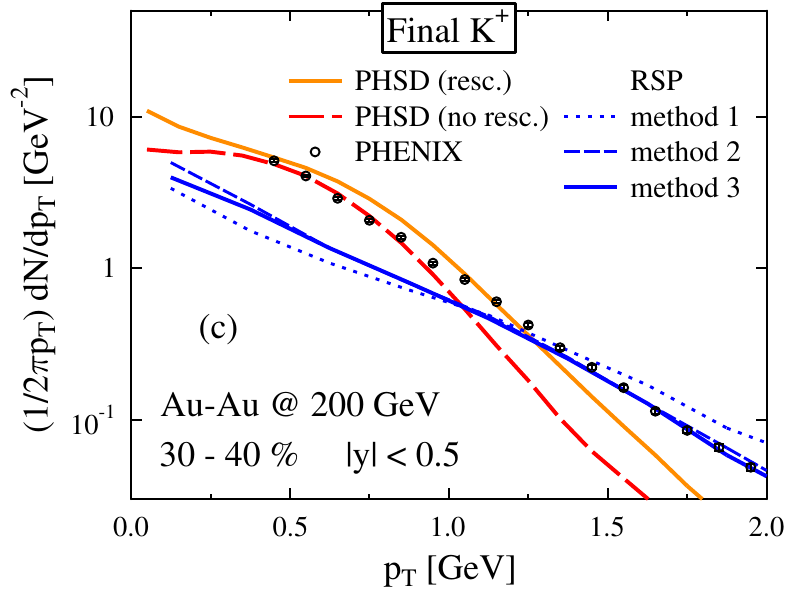} \ \ \ \
    \includegraphics[width=8cm]{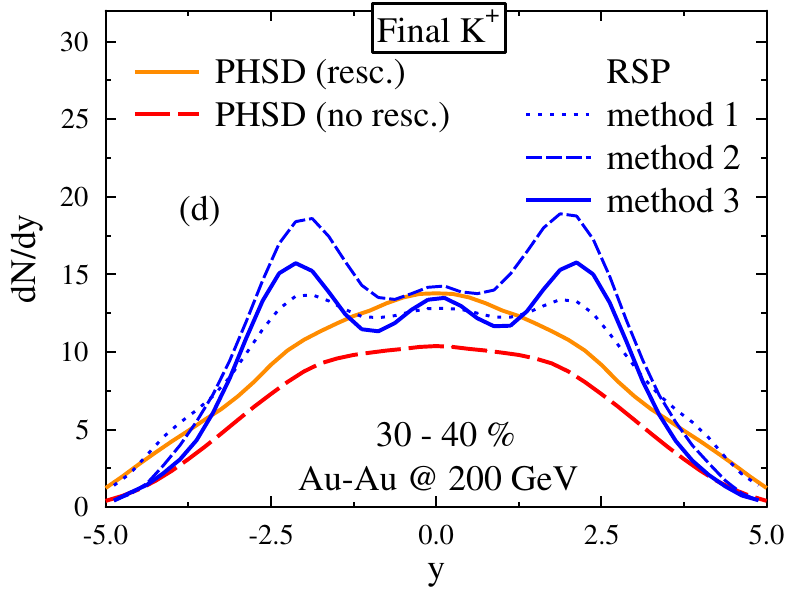}
  \end{center}
  \vskip -4.5mm
  \caption{(Color online) Final distributions of the mesons in PHSD and RSP, using the three difference conversion methods explained in the text. We display the transverse momentum spectrum for $|y| < 0.5$ and the rapidity distribution, for pions in the top panels (a) and (b), for kaons in the bottom panels (c) and (d). The PHSD results are shown with and without final state hadron rescattering. \label{results_final2}}
  \vskip -3mm
\end{figure*}

Figure \ref{results_final2} shows the corresponding final spectra of hadrons. Here $\pi$ denotes the average over $\pi^+$ and $\pi^-$ and $K$ denotes the average over $K^+$ and $K^-$.

Initially, the PHSD transverse momentum distribution of partons has a rather large slope (as compared to the final distribution). This is expected  because their mass is very large as compared to that of pions and kaons. During the expansion this large mass is converted into kinetic energy. In addition, the potential between the quarks is repulsive in PHSD which helps to flatten the slope during the expansion. Both effects are at the place where finally the PHSD spectrum agrees with the experimental spectrum up to $p_T = 1.5 \ GeV$. Above this value, jets are playing a role which is not included in the version of PHSD employed here.

The first conversion method of the PHSD initial condition gives a slope that is much flatter than the experimental data. This is a consequence of the higher momentum owing to the lower mass in RSP as compared to PHSD. Naturally, for method 2, in which the conversion creates more particles with a smaller momentum the slope is stiffer as in method 1. Finally, method 3 shifts the PHSD initial condition to larger momenta but with the same slope at high momenta as the PHSD initial condition.

In spite of the different initial transverse momentum distributions for quarks [cf. Fig. \ref{results_initial1}(a)] the final pion spectra are  relatively similar [Fig. \ref{results_final2}(a)]. In RSP the partons gain mass during the expansion and the interaction among partons is attractive. Both effects together produce an increase of the spectral slope during the expansion. For (non equilibrium) methods 2 and 3 we obtain  almost  identical pion $p_T$ spectra which are close to the PHSD pion spectra as well as to the data. Only (equilibrium) conversion method 1, which induces a very hard initial transverse momentum distribution --contrary to the initial momentum distribution in the PHSD cells-- fails to reproduce the PHSD transverse momentum spectra and the experimental data, respectively. Thus, the final pion transverse distribution can provide a constraint on the momentum distribution of the degrees of freedom in the very early phase.

For the strange quarks the initial distributions show a similar difference in the slopes as for the nonstrange quarks [cf. Fig. \ref{results_initial1}(c) vs Fig. \ref{results_final2}(c)]. There are a lot of hard collisions (following the NJL cross sections) between $s$ quarks during the expansion of the plasma, which make this transverse spectrum flatter than the initial one, despite of the attractive potential. At the end of the expansion we see that the RSP matches the high-momentum part of the experimental spectra, whereas there are too few kaons at low momentum.

This is a consequence of the strong $s \bar s \to u\bar u (d \bar d)$ cross section in NJL \cite{Rehberg1996b}, which depopulates the strange quark spectra at low-$p_T$ during the expansion and leads to a suppression of low $p_T$ kaons as compared to a calculation without this cross section. The inverse reaction is highly suppressed owing to the much larger mass of the strange quarks as compared to the light quarks. The corresponding cross section in PHSD is at least an order of magnitude smaller \cite{Ozvenchuck2013}. Because in the LUND string fragmentation the ratio of $u:d:s$, chosen as $3:3:1$ in PHSD at this energy, is a free parameter, another choice of this ratio would bring RSP closer to the data. We see here as well that methods 2 and 3 yield very similar distributions of the final kaons.

The initial parton rapidity distribution in the PHSD is peaked at midrapidity, whereas all conversion methods to RSP show for the light quarks a local minimum at midrapidity [Fig. \ref{results_initial1}(b)]. This is a consequence of the Lorentz transformation between the local rest frame and the computational frame. For the heavier strange quarks this effect is less pronounced but, being lighter than the PHSD partons, it is still present [Fig. \ref{results_initial1}(d)]. During the expansion the midrapidity region is depopulated in PHSD owing to the repulsive force between the partons, whereas in RSP the attractive force increases the midrapidity yield [Figs. \ref{results_final2}(b) and \ref{results_final2}(d)]. For the kaons this effect is more than counterbalanced owing to the strong $s \bar s \to u\bar u (d \bar d)$ cross section for small $\sqrt{s}$. Most of the difference in the midrapidity yield between RSP and PHSD comes from particles with very low $p_T$. Thus, the final rapidity distribution of hadrons --taken relative to the rapidity distribution from $p$-$p$ collisions-- provides information on the sign of the potential and interaction strength  in the QGP phase.

\begin{figure*}
  \begin{center}
    \includegraphics[width=8cm]{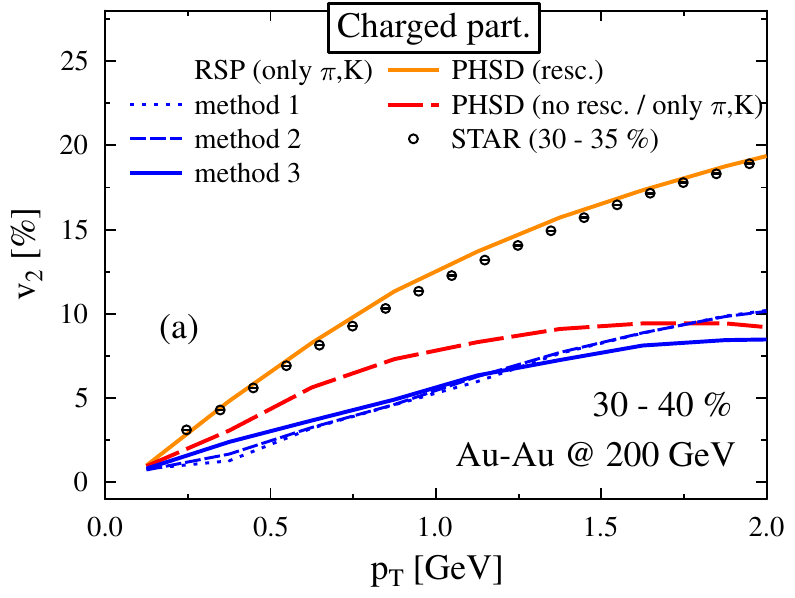} \ \ \ \
    \includegraphics[width=8cm]{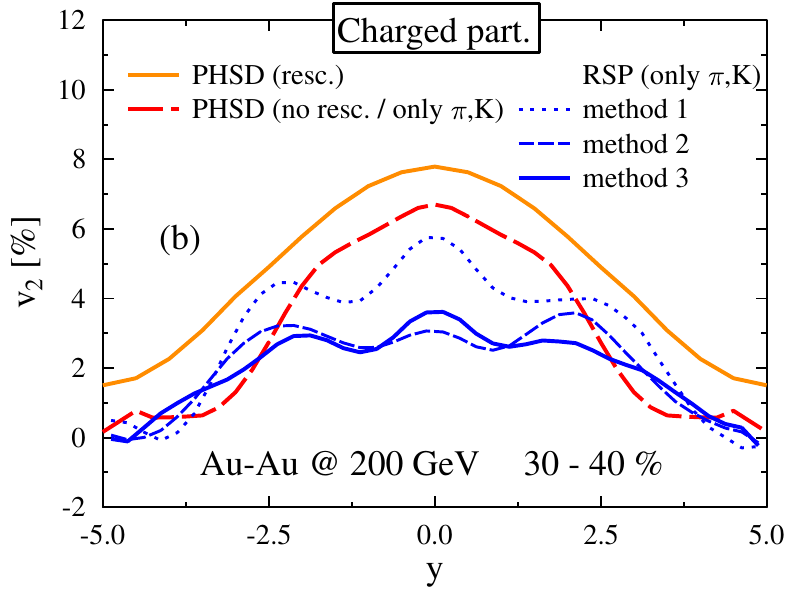}
  \end{center}
  \vskip -4.5mm
  \caption{(Color online) Elliptic flow $v_2(EP)$, as a function of transverse momentum $p_T$ (a) and of the rapidity $y$ (b) for Au--Au collisions at $\sqrt{s} = 200$ $A$GeV and $b = 8.4$ fm (30\%$-$40\% centrality) for PHSD and RSP. The different results of RSP correspond to different conversion methods of the PHSD initial condition to RSP initial conditions (see text for explanation). The PHSD calculations including hadronic rescattering are shown by full lines; the PHSD results without hadronic rescattering (and without the contribution of baryons) --to be compared to the present RSP results which do not include hadronic rescattering-- are displayed by dashed lines. \label{results_final3}}
  \vskip -2mm
\end{figure*}

\begin{figure*}
  \begin{center}
    \includegraphics[width=8.2cm]{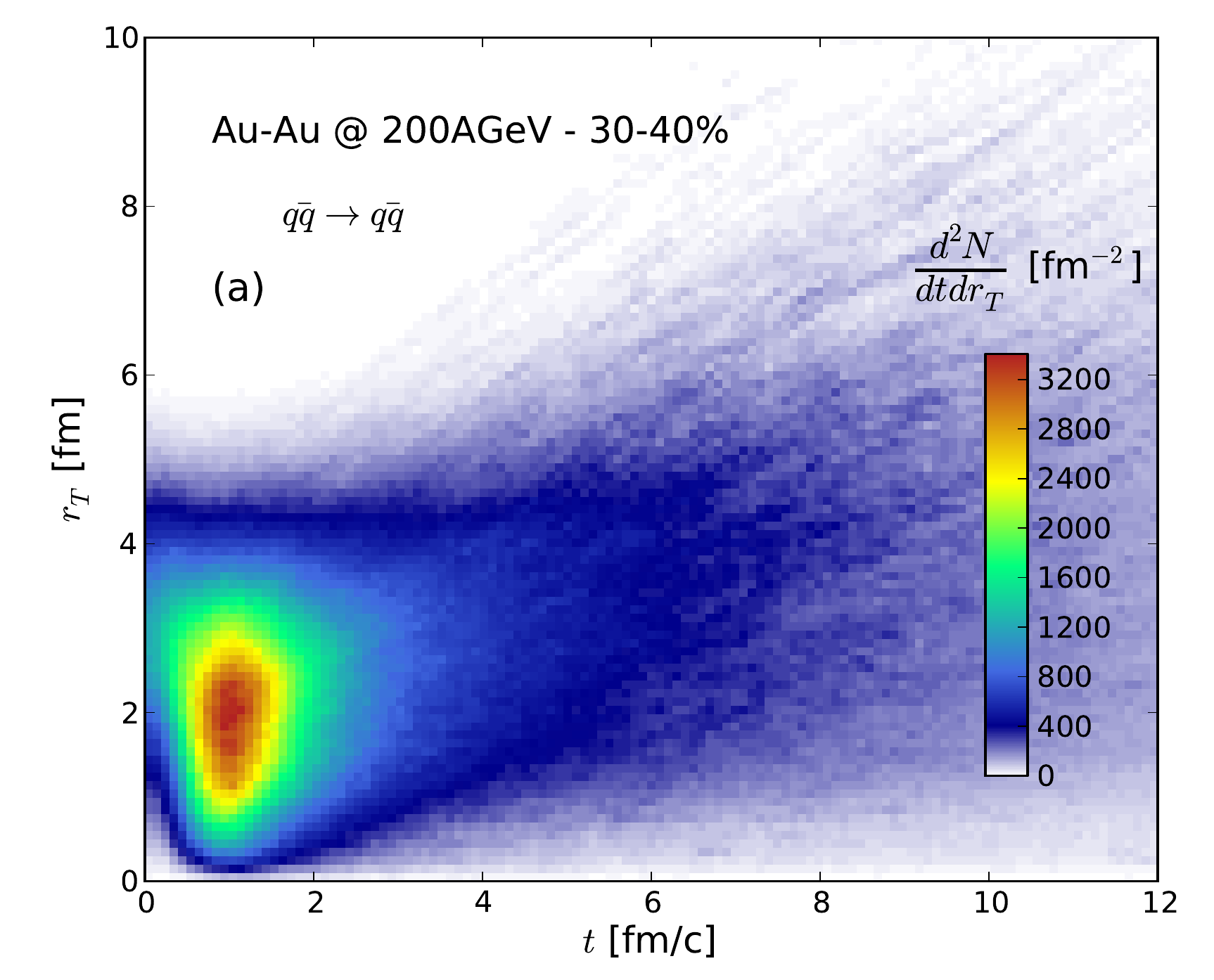} \ \
    \includegraphics[width=8.2cm]{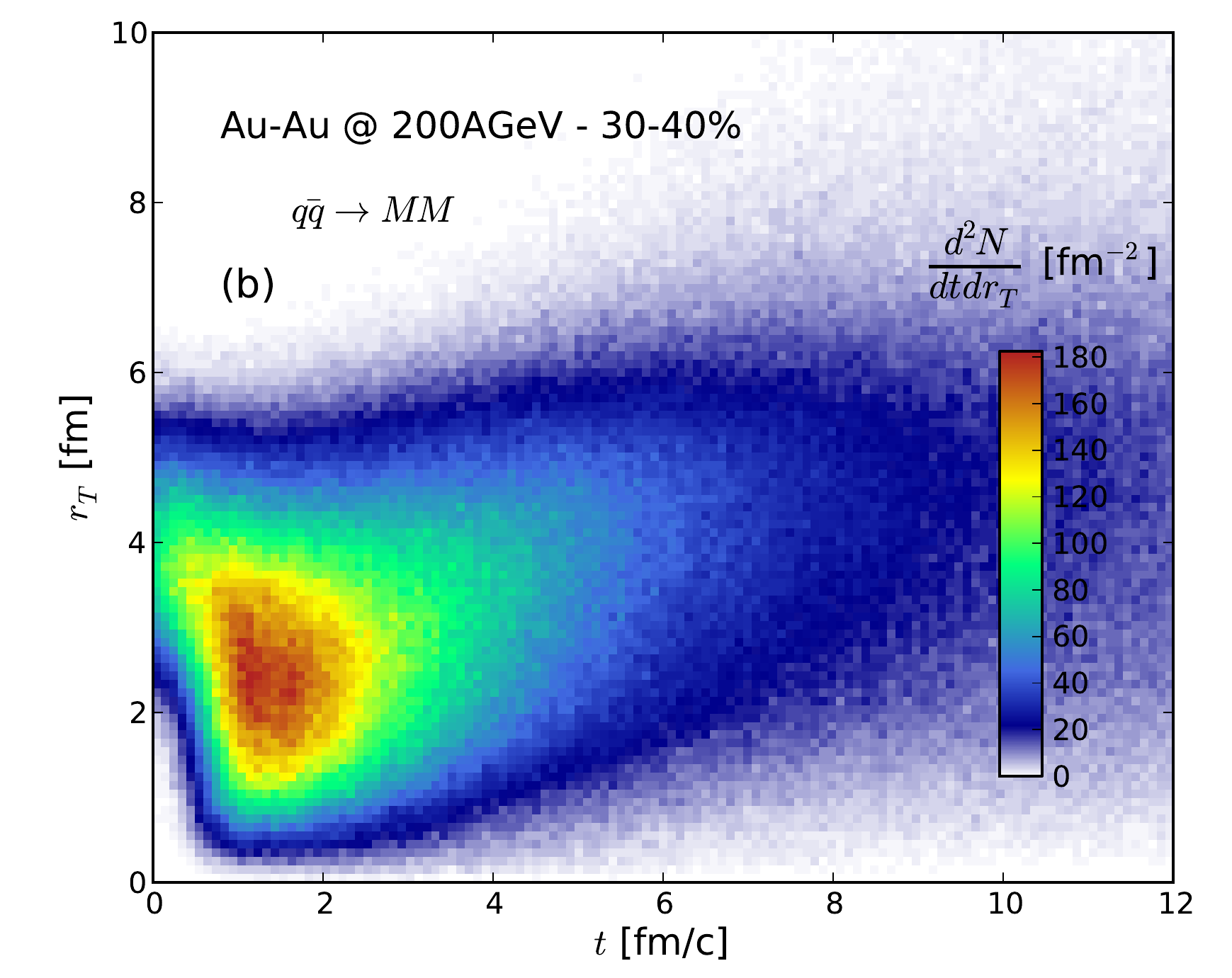}
  \end{center}
  \vskip -4mm
  \caption{(Color online) Space-time distribution $d^2 N / dt dr_T$ of elastic ($q\bar q \ \to \ q\bar q$) (a) and inelastic ($q\bar q \ \to \ MM$) (b) collisions in RSP.\label{coll_dist}}
  \vskip -3.5mm
\end{figure*}

Another observable of interest is the elliptic flow of hadrons as a function of $p_T$ and  rapidity $y$. In hydrodynamical calculations it is a direct consequence of the initial spatial eccentricity of the overlap region between projectile and target. Our results are presented in Fig. \ref{results_final3} and compared with STAR data \cite{Star2003} for the same event class. We display the PHSD results, which reproduce nicely the experimental elliptic flow if the hadronic rescattering is included. The PHSD results without hadronic rescattering and without the baryon contribution (dashed line) shows that hadronic rescattering increases $v_2$ by more than 50\%. For all three conversion methods RSP (which does not include hadronic rescattering) underpredicts the corresponding elliptic flow of PHSD. The attractive potential between the quarks in RSP counterbalances partially the transverse pressure which generates the elliptic flow. Again the rapidity distribution  $v_2(y)$ (right-hand side of Fig. \ref{results_final3}) shows more clearly the differences between the attractive and repulsive potentials in RSP and PHSD, respectively, as well as the effect of hadronic rescattering.



\vspace*{-5mm}
\section{Conclusion}
\vspace*{-3mm}

In this paper we studied the expansion of a QGP (only $q$ and $\bar q$ in RSP) and its hadronization within two different transport theories, PHSD and RSP. PHSD is based on the DQPM which reproduces the lattice equation of state, whereas RSP is based on the NJL Lagrangian, which makes it possible to calculate parton masses and parton cross section at finite temperature and quark chemical potential $\mu$. PHSD is a relativistic off-shell transport approach (incorporating resummed two-body correlations) and RSP a relativistic $n$-body theory. The parton masses in both approaches are very different. In the DQPM the partons have masses of several hundred MeV and the masses increase with temperature owing to the partonic interactions with the plasma constituents as in hard thermal loop (HTL) approaches; in the NJL model the partons in the high-temperature plasma reduce to their bare mass of a couple of MeV for the light quarks. The interactions between partons is repulsive in PHSD, whereas, taking only into account the pseudoscalar mesons, it is attractive in the NJL approach. RSP as well as PHSD are transport theories which do not require, in contrast to hydrodynamical approaches, that the system is locally in thermal equilibrium or close to it.

For our studies we employ the standard PHSD initial condition, i.e., the phase-space configuration after all initial hadrons from the Lund fragmentation are converted into quarks and gluons (the few corona particles which do not take part in the QGP are discarded). Respecting the requirement that the local energy density is conserved, we proposed three different methods to convert the PHSD initial condition into a RSP initial condition, which yield quite different RSP initial transverse momentum and rapidity distributions.

We have found that for the two very different transport theories one obtains for the same initial local energy-density distributions a relatively similar transverse momentum distributions of pions as well as of the elliptic flow $v_2$ if the non equilibrium momentum distribution of the PHSD initial condition is taken into account (methods 2 and 3). An isotropic momentum distribution in the initial cells is found to be incompatible with the PHSD results and data (method 1). The difference observed in the kaon spectrum is attributed to the large $s \bar s \to u \bar u$ cross section in the NJL approach, whereas in PHSD it is smaller and determined by the coupling strength and the gluon propagator, i.e., the gluon pole mass and width.

These findings also hold true for the rapidity distribution of kaons and pions; however, the methods 2, 3, and 1 for pions differ substantially because the Lorentz transformations from the local cell to the computational frame are different owing to the very different masses of the light quarks in PHSD and RSP. Accordingly, the rapidity distributions turn out to be more sensitive to the non-equilibrium nature of the initial conditions than the $p_T$ spectra at midrapidity because there is a partial equilibration during the dynamical expansion.

Also, the hadron $v_2$ as a function of $p_T$ does not differ substantially between PHSD (without hadronic rescattering) and RSP: While starting from zero initial $v_2$ in the both models, the elliptic flow develops practically to the same level owing to the attractive potential and rescattering of very light NJL quarks in the RSP and by the repulsive potential and more modest rescattering of heavy quarks and gluons in the PHSD. In PHSD as well as in hydrodynamical calculations the hadronic rescattering adds more than 30\% to the pion $v_2$ \cite{Werner2014}. Again the rapidity distribution of $v_2(y)$ shows a stronger sensitivity to the nature of the degrees of freedom and their interaction strength than the $p_T$ dependence of the $v_2$.

\begin{figure} [t]
  \begin{center}
    \includegraphics[width=8.5cm]{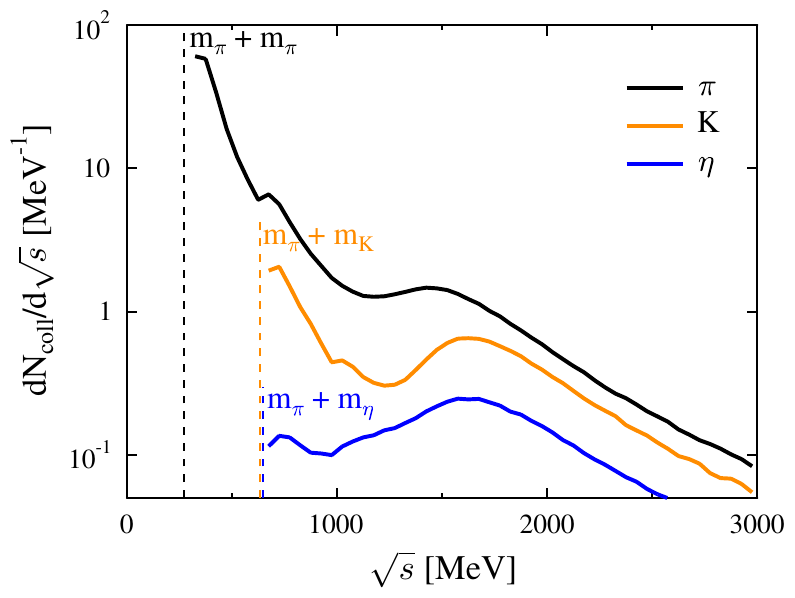}
  \end{center}
  \vskip -5mm
  \caption{(Color online) Probability distribution of hadronization collisions according to $\sqrt{s}$ from RSP.\label{collision_s_dist}}
  \vskip -4mm
\end{figure}

Thus, our study has shown that ``bulk'' observables such as transverse momentum ($p_T$) and rapidity ($y$) distributions of hadron yields as well as the elliptic flow $v_2$ reflect some traces of the non-equilibrium origin of the initial stage of the reaction. However, the partonic interactions in the QGP --which are realized very differently in the PHSD and RSP-- lead to a partial thermalization of the degrees of freedom at midrapidity which does not allow robust conclusions on the initial state configurations to be drawn.

\vspace*{-5mm}
\section*{Acknowledgment}
\vspace*{-3mm}

RM thanks H. Berrehrah for fruitful discussions and also for his continuous interest. JA thanks V. Ozvenchuck and K. Werner for discussion about the partonic cross sections and the elliptic flow. This work was supported by the ``HIC for FAIR'' framework of the ``LOEWE'' program and by the European Network Turic. The computational resources were provided by the LOEWE-CSC. This paper was supported by the project ``Together'' of the region pays de la loire.

\section{Appendix}

\subsection{The hadronization problem}

In RSP the collisions are done using the relativistic geometrical method \cite{Kodama1984} to obtain the best accuracy. We can use this method because we only consider $2\to2$ processes (no higher-order cases like $2\to3$ for instance).  In Fig. \ref{coll_dist} we show the distribution in time and space for elastic collisions (left) and for hadronization (right). One can see the quasi-free expansion at the early beginning because the cross sections are small and the hadronization, in the corona first, and then moving to the center. So we do not have a simple freeze-out surface like the Cooper-Frye method in hydrodynamical calculations.

An unavoidable problem are high-momentum partons. In the NJL model, the hadronization cross sections decrease strongly with increasing $\sqrt{s}$. Therefore, partons with large momentum $p$ (jetlike particles) cannot be hadronized except by fragmentation or $2 \to$ n processes, which is currently not included into the RSP. Such processes (like $2 \to 3$) show indeed an increase in cross sections for larger $\sqrt{s}$ \cite{Isert1998}. Figure \ref{collision_s_dist} shows the distribution of hadronization collisions as a function of $\sqrt{s}$. Hadronization is most important close to threshold. At high energy we see a broad distribution in $\sqrt{s}$ with a maximum around $\sqrt{s} = 1.6$ GeV.

\begin{figure} [t]
  \vskip  4mm
  \begin{center}
    \includegraphics[width=6.8cm]{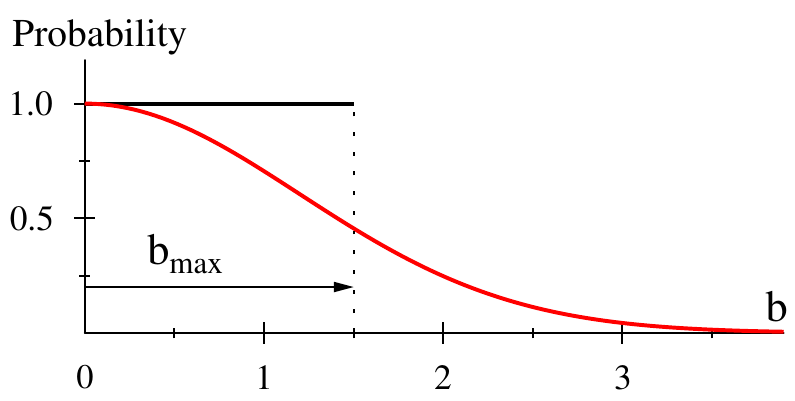}
  \end{center}
  \vskip -4mm
  \caption{(Color online) Example of a collision probability profile as a function of the the impact parameter $b$.\label{geom_cross}}
  \vskip -3mm
\end{figure}

To improve our efficiency, we have modified the standard ``billiard ball''-type (collision takes only place if $b < \sqrt{\sigma/\pi}=b_{max}$) description  by introducing a probability which depends on the distance between the particles (see Fig. \ref{geom_cross}). This is  similar to the method used in GLISSANDO 2 \cite{Rybczynski2013} for wounded nucleons. The total cross section remains the same:
\begin{equation}
  b_{\text{max}} = \sqrt{\frac{\sigma}{\pi}} = \int_0^{b_{\text{max}}} db = \int_0^\infty db \ P(b).
\end{equation}
We use $P(b) = \exp[-\frac{\pi}{4} (b / b_{\text{max}})^2]$. This method improves the hadronization for particles, especially at late times  when $b_{max}$ is of the order of the mean interparticle distance.

The few remaining quarks are finally hadronized by combining partons which are close in phase space, similar to the method of Ref. \cite{Dorso1992}.

  \bibliography{biblio}

\end{document}